\documentclass[reprint,floatfix,superscriptaddress,aps,prd,nofootinbib]{revtex4-2}

\usepackage{graphicx}
\usepackage{amsmath}
\usepackage{amssymb}
\usepackage{mathrsfs}
\usepackage{makecell} % for Xhline
\usepackage[hidelinks]{hyperref}
\usepackage[normalem]{ulem} % for strikethrough
\usepackage[usenames,dvipsnames]{color}
\usepackage{orcidlink}

\allowdisplaybreaks

\definecolor{forestgreen}{rgb}{0.2, 0.6, 0.2}

\newcommand{\Cornell}{\affiliation{Cornell Center for Astrophysics and
    Planetary Science, Cornell University, Ithaca, New York 14853, USA}}
\newcommand{\Caltech}{\affiliation{Theoretical Astrophysics 350-17,
    California Institute of Technology, Pasadena, California 91125, USA}}
\newcommand{\AEI}{\affiliation{
    Max Planck Institute for Gravitational Physics (Albert Einstein Institute), D-14476 Potsdam, Germany}}

\begin{document}

\preprint{APS/123-QED}

\title{Extending Gravitational Wave Extraction Using Weyl Characteristic Fields}
\author{Dante A. B. Iozzo
\orcidlink{0000-0002-7244-1900}}
\email{dai32@cornell.edu}
\Cornell
\author{Michael Boyle
\orcidlink{0000-0002-5075-5116}}
\Cornell
\author{Nils Deppe
\orcidlink{0000-0003-4557-4115}}
\Cornell
\Caltech
\author{Jordan Moxon
\orcidlink{0000-0001-9891-8677}}
\Caltech
\author{Mark A. Scheel}
\Caltech
\author{Lawrence E. Kidder
\orcidlink{0000-0001-5392-7342}}
\Cornell
\author{Harald P. Pfeiffer
\orcidlink{0000-0001-9288-519X}}
\AEI
\author{Saul A. Teukolsky
\orcidlink{0000-0001-9765-4526}}
\Cornell
\Caltech

\date{\today}

\begin{abstract}
  We present a detailed methodology for extracting the full set of
  Newman-Penrose Weyl scalars from numerically generated spacetimes
  without requiring a tetrad that is completely orthonormal or perfectly
  aligned to the principal null directions. We also describe how to implement
  an extrapolation technique for computing the Weyl scalars' contribution
  at asymptotic null infinity in postprocessing. These methods have been
  used to produce $\Psi_4$ and $h$ waveforms for the Simulating
  eXtreme Spacetimes (SXS) waveform catalog and now have been expanded to
  produce the entire set of Weyl scalars. These new waveform quantities are
  critical for the future of gravitational wave astronomy in order to
  understand the finite-amplitude gauge differences that can occur in numerical
  waveforms. We also present a new analysis of the accuracy of waveforms produced
  by the Spectral Einstein Code. While ultimately we expect Cauchy characteristic
  extraction to yield
  more accurate waveforms, the extraction techniques described here are
  far easier to implement and have already proven to be a viable way to produce
  production-level waveforms that can meet the demands of current gravitational-wave
  detectors.
\end{abstract}

\maketitle

%\tableofcontents

\section{Introduction}

As the field of gravitational-wave astronomy is preparing for the next
generation of detectors, it is becoming increasingly important for numerical
waveform models to achieve the high degree of accuracy that will be needed~\cite{Purrer2019}. Part
of this improvement must come from a systematic understanding of gauge effects
inherent in the waveforms. Gauge effects in a waveform, if erroneously interpreted as
physical effects, can have a direct and adverse impact on parameter
estimation from a detected gravitational wave~\cite{Boyle2016,
  Woodford2019, Kelly2013, Lehner2007}. Both phenomenological and
surrogate waveform models depend heavily on numerical relativity (NR) for
their construction~\cite{Khan2019, Field2014}. Thus we need a thorough understanding of waveform extraction
and gauge effects in numerically generated spacetimes.

Even ``gauge-invariant'' waveforms are only invariant in a very limited
technical sense of perturbation theory; such waveforms still generally have an
infinite-dimensional set of gauge freedoms described by the
Bondi-Metzner-Sachs (BMS) group---which includes the usual Poincar\'e group
along with the more general ``supertranslations''~\cite{Bondi1962, Sachs1962,
Sachs1962_2}.  These gauge freedoms are not restricted to infinitesimal transformations
and can result in appreciable finite-amplitude gauge differences especially in numerical
relativity. The BMS group induces a fractional change in the waveforms directly proportional
to the size of the gauge change.
And because gauge
conditions used in NR simulations are complicated and widely varied, we can
expect to find significant effects in waveforms due to gauge choices.  In
addition to the standard time offset and rotation gauges, important effects
due to boost and translation have already been found in numerical
waveforms~\cite{Boyle2016, Woodford2019, Kelly2013, Gualtieri2008}.  To move
beyond this basic analysis---and to understand supertranslations---we need
more information than is currently produced by most NR codes.

Often, only the gravitational wave strain $h$ and the Weyl scalar
$\Psi_4$ are extracted from NR simulations for the purpose of
constructing waveforms. While these are the quantities most directly relevant
for gravitational-wave detectors, they by no means provide complete
information regarding the curvature or radiation of the spacetime.
Since $\Psi_4$ and $h$ are particular components of tensors, we need
complete knowledge of all the components to apply transformations~\cite{Moreschi1986}.
Any attempt at understanding these gauge
freedoms and comparing different waveforms in a meaningful way requires
understanding how waveforms
behave under transformation, therefore requiring more information than $h$ and
$\Psi_4$ alone.

Accessing the information about spacetime curvature in an NR
simulation requires extracting the information stored in the Weyl tensor.
The usual prescription is to compute five
complex scalar fields from the inner product of the Weyl tensor with the
orthonormal basis vectors of a complex null tetrad~\cite{Newman1962,Hinder2011,Campanelli2006,Gunnarsen1995}.
The null tetrad can
be chosen so that the five resulting Weyl scalars are related to quantities like
the gravitational radiation or the mass and spin of the binary system~\cite{Lopez2017}.

The values of these Weyl scalars depend on the choice of the tetrad, and
so it is critical to pick a well-suited tetrad. If one is interested in comparing
the Weyl scalars across different simulations, the tetrad needs to be consistently
chosen so that gauge effects may be understood and isolated. The choice of a
consistent and suitable tetrad for NR continues to be
explored~\cite{Beetle2005,Nerozzi2005}.
Despite the ongoing challenges, a technique for
computing the Weyl scalars has been successfully implemented and used for analysis
where detailed comparison between waveforms was not necessary~\cite{Owen2010}.

Even with a consistent and well-suited tetrad choice, a coordinate system must
be used to relate the tetrads at different points. Additionally, tetrads
are defined with respect to the coordinates in NR. The issue here is that
coordinates are subject to even more freedoms than the tetrads themselves. If the
orthonormality constraint of the tetrad can be relaxed, as will be discussed in
Sec. \ref{sec:tetrad_transformations_limit}, then this
further complicates the goal of
understanding the tetrad choice---and thus the waveform quantities---across
different spacetimes.

There are two separate arenas for analyzing the curvature and radiation quantities;
each has its own motivation and its own challenges. It is important to make this
distinction and employ a technique for extracting curvature quantities
that will be best-suited for the particular analysis of interest. In
the first case, the goal is to analyze curvature quantities at finite distances
to provide information about the Petrov
classification of the spacetime, test properties of perturbed Kerr black holes,
probe regimes of strong gravity, etc.~\cite{Owen2010,Nichols2011,Zhang2012,Nichols2012,Zhang2012_2,Hinder2011}.
In the second case, the goal is to analyze
curvature quantities extrapolated or evolved to asymptotic null infinity to
compute gravitational waveforms~\cite{Boyle2019,Healy2019,Jani2016,Kelly2011,Handmer2016,Bishop2016}.

Our extraction methodology is suited for the second case, the study of asymptotic
radiation and curvature. The primary challenge in this arena is that waveforms
computed at any finite distance in a simulation domain are not entirely free of
unwanted near-field effects or other sources of gauge pollution~\cite{Lehner2007}, e.g., from the
choice of simulation coordinates. It is therefore necessary to extrapolate or
evolve the extracted waveform out to asymptotic null infinity, using either
perturbative extraction or Cauchy characteristic extraction (CCE). At the current time, a CCE code reliable
enough for supplying production-level waveforms has been developed~\cite{spectre_code,Barkett2020,Handmer2016}. However,
uncertainties in choosing the initial data for the characteristic evolution have currently prevented its use
as the primary extraction method. We use a
perturbative extraction technique with an extrapolation procedure in postprocessing
to get the final asymptotic waveforms. By using such an extrapolation procedure,
we are able to gain further computational improvements by relaxing the requirement of
working with a completely orthonormal tetrad aligned to the principal null directions.
While one should be able to arrive at the same asymptotic waveform with either
perturbative extraction or CCE, the perturbative extraction methodology described
here is simpler to implement and can serve as a point of comparison for a CCE scheme.

Although our extraction methodology does not require a completely orthonormal tetrad
aligned with the principal null directions and thus cannot be used straightforwardly to explore curvature quantities
within the simulation domain itself, we are able to get all of the necessary
curvature quantities at asymptotic null infinity and in such a frame as to allow
for the complete fixing of gauge freedom in the waveforms~\cite{Moreschi1987,Moreschi1988}.

Our extraction methodology is based on the idea of using the characteristic
fields of the Weyl tensor evolution equations, which has already
been successfully implemented in the Spectral Einstein Code (SpEC)~\cite{SpECCode}. This
technique has proven remarkably robust and accurate by serving as the primary means of wave
extraction for the Simulating eXtreme Spacetimes (SXS) waveform catalog, the largest
catalog of numerical waveforms available~\cite{Catalog}. Previously, only $\Psi_4$
and $h$ have been extracted. For the first time, we expand this method to include
the full set of Weyl scalars and produce production-level waveforms. Using these
new quantities, we use the Bianchi identities to present a new analysis testing
SpEC waveforms against exact general relativity, providing a hard upper bound for
the accuracy of the waveforms.

\section{Extraction Methodology}

\subsection{Overview}

In order to express the ten independent components of the Weyl tensor as the
five complex Weyl scalars, we need to first define a complex
null tetrad from a linear combination of coordinate basis vectors. Consider a
4-dimensional spacetime%
\footnote{\label{fn:conventions}
  We make use of the conventions described in Appendix C
  of~\cite{Boyle2019}. Note that unlike in~\cite{Boyle2019}, the letters $(a,b,c,d,e)$
  are reserved for four-dimensional spacetime indices and the letters $(i,j,k,l,p,q)$
  are reserved for three-dimensional spatial indices.
}
described by the metric $g_{ab}$ with spherical coordinate basis vectors
$(t^a, r^a, \theta^a, \phi^a)$. We can construct linear combinations of these
basis vectors to define a complex null tetrad $(\ell^a, n^a, m^a, \bar{m}^a)$
with two real null vectors, $\ell^a$ and $n^a$, a complex vector $m^a$, and its
complex conjugate $\bar{m}^a$, such that $-\ell_a n^a = m_a \bar{m}^a = 1$ and
all other inner products vanish. The vectors $\ell^a$ and $n^a$ are aligned with
outgoing and ingoing null geodesics.

For any choice of complex null tetrad, we can define the Weyl scalars as the
inner products of the Weyl tensor and the null tetrad vectors,
\begin{subequations} \label{eq:Weyl_scalar_defs}
\begin{align}
  \Psi_4 &= C_{abcd} n^a \bar{m}^b n^c \bar{m}^d, \\
  \Psi_3 &= C_{abcd} l^a n^b \bar{m}^c n^d, \\
  \Psi_2 &= C_{abcd} l^a m^b \bar{m}^c n^d, \\
  \Psi_1 &= C_{abcd} l^a n^b l^c m^d, \\
  \Psi_0 &= C_{abcd} l^a m^b l^c m^d.
\end{align}
\end{subequations}
The Weyl scalars are thus intimately connected to the tetrad choice. It is
critically important for a consistent tetrad choice to be established for
two reasons. First, so that the Weyl scalars will most readily reveal properties
of the spacetime. Second, to enable a meaningful comparison of the Weyl scalars
across timesteps of a single simulation or across different simulations altogether.

Several approaches for constructing a tetrad in NR involve
solving for the principal null directions of the spacetime or
performing a procedure to orthonormalize a coordinate tetrad and then
effectively solving for the Lorentz transformation to achieve
a desired frame~\cite{Nerozzi2005,Nerozzi2017,Campanelli2006}.
However, for the purpose of
measuring the Weyl scalars at asymptotic null infinity $\mathscr{I}^{+}$,
we are only interested in the leading-order contributions. Accordingly, we don't
need such an involved procedure and can take advantage of a perturbative extraction
technique in which any errors become negligible at $\mathscr{I}^{+}$.

This technique is simpler to formulate, easier to implement, and results in asymptotic
waveforms that are
of primary interest for gravitational wave astronomy.
These waveforms are still subject to the infinite-dimensional gauge freedom
at $\mathscr{I}^{+}$, described by the Bondi-Metzner-Sachs (BMS) group.
However, in principle a method exists for partially fixing the BMS gauge freedoms of
the waveforms obtained from our perturbative extraction
technique~\cite{Moreschi1987, Moreschi1988}.

In addition to the Weyl scalars, we extract the gravitational wave strain $h$
from the simulations via the Regge-Wheeler-Zerilli (RWZ) extraction procedure~\cite{Sarbach2001,Regge1957,Zerilli1970}.

\subsection{Weyl Characteristic Fields}

The goal of this section is to write the Weyl scalars in terms of the various
characteristic fields of the Weyl tensor evolution equation instead of in terms
of the full four-dimensional Weyl tensor itself~\cite{Gunnarsen1995,Kidder2005,alcubierre2008,Pratten2015}.
Numerical relativists typically employ a
3+1 decomposition of the spacetime, foliating the spacetime by spatial
hypersurfaces $\Sigma$~\cite{Arnowitt1959,baumgarte2010}.
Following this procedure,
we would want to express the Weyl
tensor in terms of quantities that can be computed on each spatial
hypersurface.
More specifically, by doing this we wish to minimize the amount of numerical noise
introduced when computing the full spacetime Weyl tensor, which normally requires
second derivatives of the spacetime metric.

For a timelike unit vector field $s^a$ orthogonal to $\Sigma$, we can define the
induced metric $\gamma_{ab} = g_{ab} + s_a s_b$. We also introduce the lapse
function $N$ and shift vector $N^a$,
\begin{subequations}
\begin{align}
  N   &= - t^a s_a,\\
  N_a &= \gamma_{ab} t^b,
\end{align}
\end{subequations}
to relate how the coordinates on $\Sigma$ evolve from one hypersurface to the next.
By solving the generalized harmonic formulation of the Einstein equations, we
compute not only $\gamma_{ab}$ but also its first derivatives
$d_{abc} \equiv \partial_a \gamma_{bc}$ algebraically from the evolved
variables~\cite{Kidder2001}. Thus, we can compute the spatial components of the extrinsic
curvature as
\begin{equation}
  K_{ij} = - \frac{1}{2N} \left ( d_{0ij} - N^k d_{kij} - 2 \gamma_{k(i} \partial_{j)} N^k \right ).
\end{equation}
Following Eq.~$(2.20)$ in~\cite{Kidder2001}, we can also compute the spatial
components of the spatial Ricci tensor $R_{ij}$ from $d_{ijk}$ and
$\partial_i d_{jkl}$. All the terms involved in computing $K_{ij}$ and $R_{ij}$
can either be taken directly from the evolved variables or require an additional
\textit{spatial} derivative. By using a pseudospectral code, spatial derivatives
of quantities on $\Sigma$ can be computed by spectral differentiation instead of
having to use a finite-difference method. However,
we cannot compute the full Weyl
tensor using quantities available on $\Sigma$, so we
must proceed to project the
curvature information of the Weyl tensor onto $\Sigma$.

Having already chosen a timelike unit vector field $s^a$ orthogonal to $\Sigma$,
the Weyl tensor can be split into an electric and magnetic part,
\begin{subequations}\label{eq:weyl_electric_magnetic}
\begin{align}
  E_{ij} &= C_{acbd} s^a s^b \gamma^{c}{}_{i} \gamma^{d}{}_{j}, \\
  B_{ij} &= -C_{acbd}^\star s^a s^b \gamma^{c}{}_{i} \gamma^{d}{}_{j},
\end{align}
\end{subequations}
where $C_{abcd}^\star = \frac{1}{2}  C_{abef}\epsilon^{ef}_{\hphantom{ef}cd}$ is the
right dual of the Weyl tensor.  These tensors are symmetric, traceless, and
orthogonal to $s^a$. The electric Weyl tensor $E_{ij}$ is the tidal tensor on
the spatial hypersurface, and the magnetic Weyl tensor $B_{ij}$ encodes the
differential frame-dragging on the spatial hypersurface~\cite{Nichols2011,Zhang2012,Nichols2012}.
What makes this approach particularly attractive in NR is that $E_{ij}$ and
$B_{ij}$ can be computed directly from $K_{ij}$ and $R_{ij}$~\cite{Gunnarsen1995,alcubierre2008},
\begin{subequations}
\begin{align}
  E_{ij} &= R_{ij} + K^{k}{}_{k} K_{ij} - K_{i}{}^{k} K_{kj}, \label{eq:E_canonical} \\
  B_{ij} &= \mathcal{D}_k K_{l(i} \epsilon_{j)}{}^{lk}, \label{eq:B_canonical}
\end{align}
\end{subequations}
where $\epsilon_{ijk}$ is the Levi-Civita tensor on $\Sigma$, $\mathcal{D}_i$ is
the spatial covariant derivative on $\Sigma$, and nonvacuum terms have been omitted.

Comparing $E_{ij}$ and $B_{ij}$ in Eqs.~\eqref{eq:weyl_electric_magnetic} with how
the Maxwell electric and magnetic fields are defined on a spatial hypersurface,
it is clear that the Weyl tensor is taking the place of the Faraday
tensor~\cite{Maartens2008,Maartens1998,Novello1980}. We can continue with this
mathematical analogy to develop a pair of coupled evolution equations for $E_{ij}$
and $B_{ij}$ that bear a strong resemblance to the Maxwell equations~\cite{Maartens2008}.
For details, see Appendix~\ref{app:complex_char_fields}. The six real-valued
characteristic fields of these coupled evolution equations are given by
\begin{subequations}\label{eq:characteristic_fields}
\begin{align}
  U^{\pm}_{ij} &= \left ( E_{kl} \pm \epsilon_{k}{}^{pq} r_q B_{lp}\right ) \left ( q^{k}{}_{i} q^{l}{}_{j} + \frac{1}{2} q^{kl} q_{ij} \right ), \\
  V^{\pm}_{i} &=  \left ( E_{kl} \pm \epsilon_{k}{}^{pq} r_q B_{lp}\right ) r^l q^{k}{}_{i}, \\
  \mathcal{E} &= E_{ij} r^i r^j, \label{eq:tendicity}\\
  \mathcal{B} &= B_{ij} r^i r^j, \label{eq:vorticity}
\end{align}
\end{subequations}
where $r^i$ is the radial unit vector on $\Sigma$ and $q_{ij}$ is the spatial
2-sphere metric orthogonal to $r^i$. The $U^{\pm}_{ij}$ tensors, being spatial,
symmetric, and transverse-traceless, have two independent components that describe
the two gravitational wave degrees of freedom. The fields $\mathcal{E}$ and
$\mathcal{B}$ are the tendicity and vorticity of the spatial
hypersurface~\cite{Nichols2011, Zhang2012, Nichols2012}.

All that remains is to specify the tangent vectors $\theta^a$ and $\phi^a$ on the
2-sphere orthogonal to $r^a$ and then construct the complex null vector $m^a$,
\begin{equation}
  m^a = \frac{1}{\sqrt{2}r} \left ( \theta^a + \frac{i}{\sin\theta} \phi^a \right ),
\end{equation}
where $r$ is the coordinate radius. This allows us to write the Weyl scalars as
inner products of the characteristic fields with $m^a$ and $\bar{m}^a$,
\begin{subequations}\label{eq:weyl_from_char_fields}
\begin{align}
    \Psi_4 &= U^{+}_{ij} \bar{m}^i \bar{m}^j, \\
    \Psi_3 &= \frac{1}{\sqrt{2}} V^{+}_{i} \bar{m}^i, \\
    \Psi_2 &= \frac{1}{2} \left ( \mathcal{E} + i \mathcal{B} \right ), \\
    \Psi_1 &= - \frac{1}{\sqrt{2}} V^{-}_{i} m^i, \\
    \Psi_0 &= U^{-}_{ij} m^i m^j.
\end{align}
\end{subequations}

\subsection{Tetrad Transformations in the Asymptotic Limit}
\label{sec:tetrad_transformations_limit}

In addition to any computational improvement that results from computing the
Weyl scalars from the real characteristic fields of the Weyl tensor evolution
system, a further improvement can be made by relaxing two requirements of our
tetrad. First, a completely orthonormal tetrad is not
necessary, and second,
the $\ell^a$ and $n^a$ vectors need not be aligned to the outgoing and ingoing
null geodesics. Even a Schwarzschild spacetime with the center of mass shifted
from the coordinate center will result in our $\ell^a$ and $n^a$ being
misaligned. Handling $\ell^a$ and $n^a$ misaligned with outgoing and ingoing
geodesics will be discussed in Sec.~\ref{sec:extrapolation}.

The reason the tetrad is not orthonormal for general spacetimes is that we
define $\theta^a$ and $\phi^a$ in Cartesian coordinates for a sphere in flat spacetime.
Using the flat spacetime $\theta^a$ and $\phi^a$ has the advantage that $m^a$ can be computed once at the
start of the simulation and then cached.
 We demonstrate that these limitations can be safely
mitigated for specific applications of the extraction procedure.

Ultimately, we are interested in the asymptotic Weyl scalars. Our ``relaxed''
tetrad can be thought of as a transformation of an aligned, orthonormal tetrad.
Since it is the leading-order behavior of the Weyl scalars that contributes to
the asymptotic data, we ask which tetrad transformations leave the leading-order
behavior invariant in an asymptotically flat spacetime.

Consider a physical spacetime $(M, g_{ab})$ conformally related to a spacetime
$(\mathfrak{M}, \mathfrak{g}_{ab})$ such that
\begin{equation}
  \mathfrak{g}_{ab} = \Omega^2 g_{ab},
\end{equation}
where $\Omega \ge 0$ is a smooth function. The manifold $\mathfrak{M}$ has a
boundary $\mathscr{I}^{+} = S^2 \times \mathbb{R}$ that terminates all
future-directed null geodesics, with $\Omega=0$ and $d\Omega \neq 0$ at
$\mathscr{I}^{+}$. The expected asymptotic behavior of the Weyl scalars is given
by the peeling theorem,
\begin{subequations}
\begin{align}
  \Psi_4 &= \mathcal{O}(\Omega),   \\
  \Psi_3 &= \mathcal{O}(\Omega^2), \\
  \Psi_2 &= \mathcal{O}(\Omega^3), \\
  \Psi_1 &= \mathcal{O}(\Omega^4), \\
  \Psi_0 &= \mathcal{O}(\Omega^5).
\end{align}
\end{subequations}

A general tetrad transformation will introduce new terms to the Weyl scalars.
However, any new terms that are higher order in $\Omega$ than the leading Weyl
scalar term will not contribute at $\mathscr{I}^{+}$. To find which tetrad
transformations are allowed, we can relate the tetrad basis vectors
$(\ell^a, n^a, m^a, \bar{m}^a)$ on $M$ to the tetrad basis vectors
$(\mathfrak{l}^a, \mathfrak{n}^a, \mathfrak{m}^a, \bar{\mathfrak{m}}^a)$ on
$\mathfrak{M}$,
\begin{subequations} \label{eq:tetrad_conformal_relation}
\begin{align}
  \ell^a &= \Omega^2 \mathfrak{l}^a, \\
  n^a &= \mathfrak{n}^a, \\
  m^a &= \Omega \mathfrak{m}^a, \\
  \bar{m}^a &= \Omega \bar{\mathfrak{m}}^a.
\end{align}
\end{subequations}
If the tetrad on $M$ is transformed such that the new terms are subleading in
$\Omega$, then taking the limit $\Omega \rightarrow 0$ will lead to the same
asymptotic tetrad. If the asymptotic tetrad is invariant, then the asymptotic
Weyl scalars will be invariant to such a transformation. In this case,
we can expect that the Weyl scalars computed at finite radii in a
simulation domain with a nonorthonormal tetrad should converge to the
asymptotic Weyl quantities with increasing radius.

It is important to note that unlike the Weyl scalars and the strain $h$,
the Newman-Penrose shear $\sigma$ actually depends on \textit{subleading}
terms of the tetrad vectors. Therefore, the asymptotic value of $\sigma$
is still not invariant under these transformations. For a full discussion, see
Appendix~\ref{sec:abandon_all_hope_ye_who_enter_here}. We do not extract
$\sigma$ from simulations so this does not present an issue to our current
considerations.

Constructing the Weyl scalars using
Eqs.~\eqref{eq:weyl_from_char_fields} is mathematically equivalent to
contracting the following tetrad with the full spacetime Weyl tensor as in
Eqs.~\eqref{eq:Weyl_scalar_defs},
\begin{subequations} \label{eq:tetrad}
\begin{align}
    \ell^a &= \frac{1}{\sqrt{2}} \left ( s^a + r^a \right ), \label{eq:tetrad_l}\\
    n^a &= \frac{1}{\sqrt{2}} \left ( s^a - r^a \right ), \label{eq:tetrad_n}\\
    m^a &= \frac{1}{\sqrt{2}r} \left ( \theta^a + \frac{i}{\sin\theta} \phi^a \right ), \label{eq:tetrad_m} \\
    \bar{m}^a &= \frac{1}{\sqrt{2}r} \left ( \theta^a - \frac{i}{\sin\theta} \phi^a \right ). \label{eq:tetrad_mbar}
\end{align}
\end{subequations}
Our choice of $s^a$ and $r^a$ ensures that $l^a l_a = n^a n_a = 0$ and
$l^a n_a = -1$ within machine precision, even at finite radii. Furthermore,
since $\theta^a$ and $\phi^a$ are defined on spheres orthogonal to $r^a$, we
can ensure $l^a m_a = n^a m_a = 0$ as well. Even with respect to the full
spacetime metric, we find $l^a m_a = n^a m_a = 0$ within machine precision
asymptotically. At this point we choose $\theta^a$
and $\phi^a$ as defined on a sphere in flat spacetime, which implies that we
cannot guarantee $m^a m_a = 0$ or $m^a \bar{m}_a = 0$ at finite radii. If we
still expect our choice of $m^a$ to be complex null and normalized at
$\mathscr{I}^{+}$, then from Eqs.~\eqref{eq:tetrad_conformal_relation} we would
hope to find the following asymptotic behavior,
\begin{subequations}\label{eq:m_at_scri}
\begin{align}
  m^a m_a &= \mathcal{O}(\Omega), \\
  m^a \bar{m}_a &= 1+\mathcal{O}(\Omega).
\end{align}
\end{subequations}
We do in fact find this behavior even in binary black hole spacetimes using
SpEC, for which the error in Eq.~\eqref{eq:m_at_scri}
at $\mathscr{I}^+$ is typically $\mathcal{O}(10^{-8})$. Since this is below our
desired tolerance, we proceed without needing any further manipulation of $m^a$.

\section{Implementation}

\subsection{Extrapolation}
\label{sec:extrapolation}

The next problem to consider is that the accuracy of the extracted waveform is
directly related to how far away from the center of the simulation domain the extraction is performed. The typical simulation
domain extends to a coordinate radius $r \sim 10^3\,M$ at most, and even at this
radius the near-field, gauge, and tetrad effects contribute up to about 1\% of the
waveform's amplitude. If we can choose a suitable conformal scaling function $\Omega=\Omega(r)$ that
accurately models the falloff of the finite-radius data, then we can set up a
procedure to extrapolate the data along null rays to $\Omega=0$. This would then be
the asymptotic data.

There are two challenges to setting up this extrapolation procedure. The first is that
for a choice of simulation coordinates $(t,x,y,z)$, the coordinate time $t$ and coordinate
radius $r=\sqrt{x^2+y^2+z^2}$ may not parametrize a null ray simply as $u=t-r$, which means
our tetrad may be misaligned even asymptotically. This would require a more clever choice
of $u=u(t,r)$. The second challenge is defining an appropriate conformal scaling
function $\Omega=\Omega(r)$. If these two issues are addressed, then we can loosely lay out
our extrapolation procedure as:
\begin{enumerate}
  \item Extract each Weyl scalar $\Psi_n(t,r,\theta,\phi)$, for $n\in\{0,1,2,3,4\}$,
        at multiple radii each timestep.
  \item For each value of $u$, separately fit the real and imaginary parts of
        $\Psi_n$ extracted at various radii to a polynomial in $\Omega$.
  \item Take the value of the polynomial with $\Omega=0$ to find asymptotic data at
        the particular $u$, and repeat for all $u$.
\end{enumerate}
This procedure will now be described in greater detail.

We note that each Weyl scalar $\Psi_n$ can be expressed as an expansion in
powers of the conformal scaling function $\Omega$ with the leading term set by
the peeling theorem,
\begin{equation}
   \Psi_n = \Omega^{5-n} \left ( \Psi^0_n + \Psi^1_n \Omega + \Psi^2_n \Omega^2 + \mathcal{O}(\Omega^3) \right ). \label{eq:peeling}
\end{equation}
The leading coefficient $\Psi^0_n(u,\theta,\phi)$ is the asymptotic Weyl scalar
on $\mathscr{I}^{+}$, so we wish to isolate $\Psi^0_n$ from the extracted finite
radius data $\Psi_n$. Since we are primarily interested in the radial dependence
we can decompose the Weyl scalars in terms of the spin-weighted spherical
harmonics (SWSHs),
\begin{equation}\label{eq:swsh_decomposition}
  \Psi_n(t,r,\theta,\phi) = \sum_{\ell=|s|}^{\infty} \sum_{|m| \le \ell} \Psi_n^{(\ell,m)}(t,r)\, {}_{s}Y_{\ell m}(\theta, \phi),
\end{equation}
for spin weight $s=2-n$. By working with the mode weights $\Psi_n^{(\ell,m)}$, we can ignore the
angular dependence. We can compute the mode weights in the decomposition by
exploiting the orthogonality of the SWSHs,
\begin{subequations}
\begin{align}
  \Psi_4^{(\ell,m)} &= \int_{S^2} \left ( U^{+}_{ij} \bar{m}^i \bar{m}^j \right ) {}_{-2}\bar{Y}_{\ell m}\, r^2 \sin\theta\, d\theta\, d\phi, \\
  \Psi_3^{(\ell,m)} &= \frac{1}{\sqrt{2}}\int_{S^2} \left ( V^{+}_{i}  \bar{m}^i \right ) {}_{-1}\bar{Y}_{\ell m}\, r^2 \sin\theta\, d\theta\, d\phi, \\
  \Psi_2^{(\ell,m)} &= \frac{1}{2} \int_{S^2} \left( \mathcal{E} + i \mathcal{B} \right)  \bar{Y}_{\ell m}\, r^2 \sin\theta\, d\theta\, d\phi, \\
  \Psi_1^{(\ell,m)} &= -\frac{1}{\sqrt{2}}\int_{S^2} \left ( V^{-}_{i}  m^i \right ) {}_{1}\bar{Y}_{\ell m}\, r^2 \sin\theta\, d\theta\, d\phi, \\
  \Psi_0^{(\ell,m)} &= \int_{S^2} \left ( U^{-}_{ij}  m^i m^j \right ) {}_{2}\bar{Y}_{\ell m}\, r^2 \sin\theta\, d\theta\, d\phi,
\end{align}
\end{subequations}
where we have used Eqs.~\eqref{eq:weyl_from_char_fields} to write the Weyl
scalars in terms of the characteristic fields. The mode weights are computed
on a set of concentric spheres of constant coordinate radius $r$ at each time step.
To compute the mode weights up to $\ell_\text{max}$, it suffices to have
$(2\ell_\text{max}+1)^2$ points on each extraction sphere evenly spaced in $\theta$
and $\phi$. Since the complex null tetrad vector $m^a$ is defined for a sphere
in flat space, it does not change throughout the simulation. Therefore, we can
precompute most of the integrand at each $(\theta,\phi)$ point, only needing to
update the characteristic fields each time step.

\begin{figure}[t]
  \centering
  \includegraphics[width=\columnwidth]{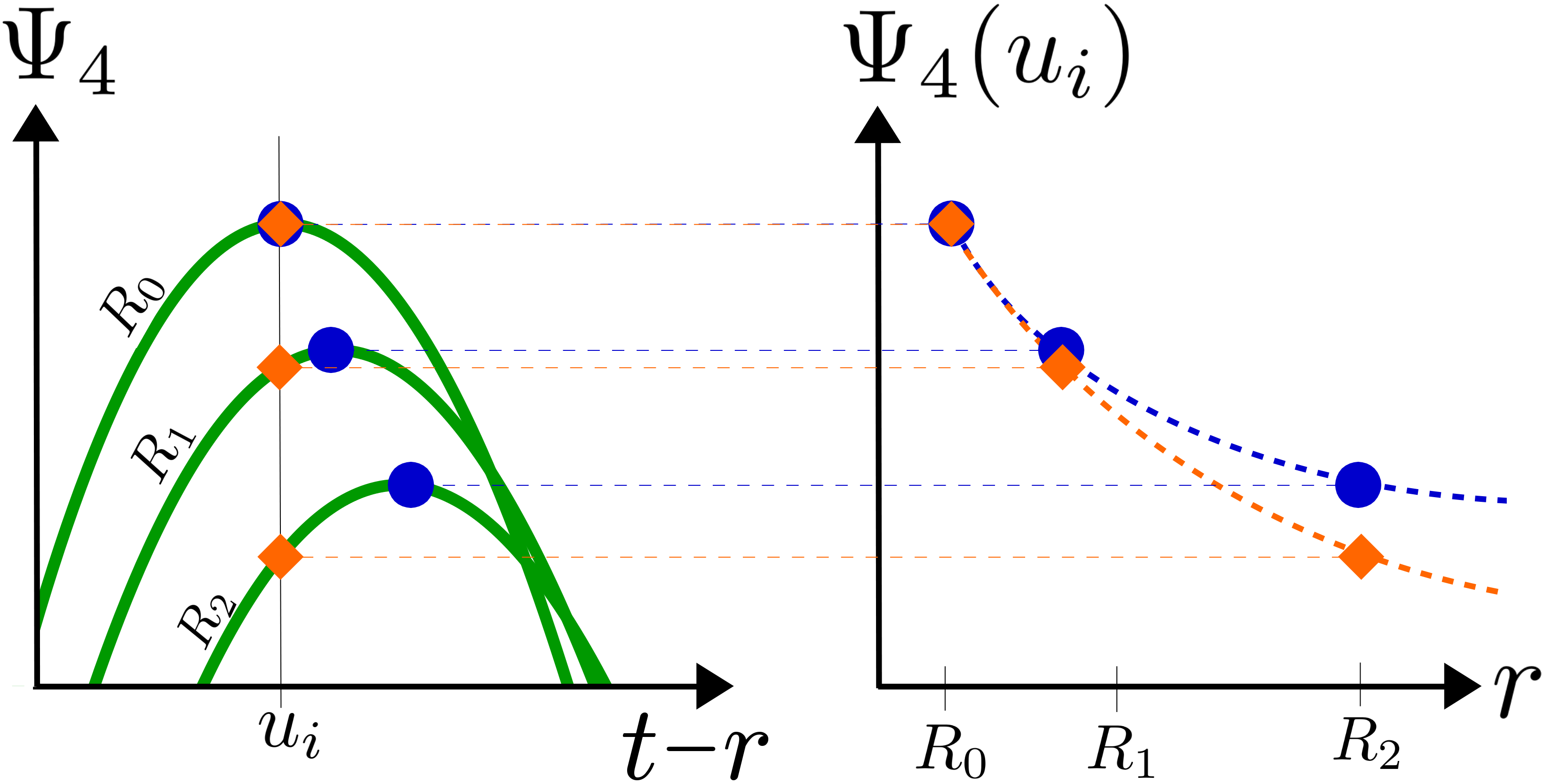}
  \caption{%
    An example of how a poor choice of retarded time $u(t,r)$ fails to capture the
    falloff behavior of gravitational radiation along an approximate outgoing null ray.
    The plot on the left illustrates a small section of three $\Psi_4$ waveforms
    extracted at different radii, one green curve per extraction radius.
    The blue dots represent points that lie along a true null ray.
    If the choice of $u(t,r)$ accurately parametrizes the null rays then the curves
    should be aligned, i.e. the blue dots should all
    be vertically aligned having the same value of $u_i$. Because of a poor choice of $u(t,r)$, in this example $u=t-r$ using
    simulation coordinates, the curves are all
    misaligned and the orange diamonds denote the values of $\Psi_4$ along the
    \textit{approximate} null ray. The right plot illustrates how the values of
    $\Psi_4$ as a function of radius along an approximate null ray (orange diamonds) deviate
    from the values along the true null ray (blue dots). Extrapolating polynomials
    are shown in the right plot as thick dashed lines for both data. Notice that
  extrapolating $\Psi_4(u_i)$ along $r\rightarrow\infty$ leads to an incorrect
  asymptotic value if a bad $u(t,r)$ is chosen.
  }
  \label{fig:wrong_u}
\end{figure}

With the angular dependence factored out of the Weyl scalars, we proceed to find
a good definition of $u(t,r)$ and $\Omega(r)$. The naive choice $u=t-r$ does not
leave us with a good parametrization of the null rays asymptotically. This is to
be expected, since the simulation coordinates were not chosen for this intent.
An attempt at improving this might be to use the radial tortoise
  coordinate $r_*$ to define the parametrization $u=t-r_*$. However, this shows
  only a marginal improvement and still does not leave us with a good enough
  parametrization for effective extrapolation.
Figure~\ref{fig:wrong_u} illustrates how a choice of $u(t,r)$ that
fails to accurately parametrize null rays will adversely affect the
extrapolation.

A better choice for a
retarded time $u$ that parametrizes outgoing null rays in the
asymptotic limit involves a radial tortoise coordinate $R_*$ constructed from the
areal radius as well as a choice of ``corrected time'' $t_\text{corr}$~\cite{Boyle2009},
\begin{subequations} \label{eq:corrected_u}
\begin{align}
  u &= t_\text{corr} - R_*, \label{eq:u_ansatz} \\
  R_* &= R + 2 M_\text{\tiny ADM} \log \left ( \frac{R}{2M_\text{\tiny ADM}} - 1 \right ), \\
  t_\text{corr} &= \int_0^T \frac{\langle N \rangle} {\sqrt{1 - 2M_\text{\tiny ADM}/R}} dT',
\end{align}
\end{subequations}
where $M_\text{ADM}$ is the ADM energy of the initial data at the start of the
simulation, $\langle N \rangle$ is the average value of the lapse over the
extraction sphere, and $R$ is the areal radius defined by computing the surface
area of the extraction sphere,
\begin{equation}
  R = \left [ \frac{1}{4\pi} \oint \sqrt{\text{det}(g_{ab})}\, d\Omega^2  \right ]^{1/2}.
\end{equation}
For the conformal scaling function, it suffices to use the inverse areal radius,
\begin{equation} \label{eq:conformal_scaling_function}
  \Omega = R^{-1}.
\end{equation}

We also scale out the waveform data's leading falloff in $R$ so that the data to be fitted is
as constant in $R$ as possible. Additionally, we scale the data by an appropriate
factor of the mass of the system $M$ so that we can work with the dimensionless quantity
$R^{5-n} M^{n-3} \, \Psi_n^{(\ell,m)}$. For our work, we choose the system mass $M$ to
be the sum of the Christodoulou masses of the black holes measured at the earliest time
after initial transients, i.e. junk radiation, have decayed from the simulation.

The goal is to find a least-squares polynomial fit in $R$ to data from a discrete
set of extraction radii $\{R_j\}=\{R_\text{min}, \dots, R_\text{max}\}$. Thus for
each mode $(\ell,m)$ and time $u_i$, there are two sets of data,
\begin{subequations}\label{eq:data_to_extrapolate}
\begin{align}
  \{\Xi_j\} &= \left \{R_{j}^{5-n} M^{n-3} \, \Re\Psi_n^{(\ell,m)}(u_i, R_j) \right \}, \\
  \{\Upsilon_j\} &= \left \{R_{j}^{5-n} M^{n-3} \, \Im\Psi_n^{(\ell,m)}(u_i, R_j) \right \},
\end{align}
\end{subequations}
to which we perform the following polynomial fit solving for coefficients
$\xi^{(i)}$ and $\zeta^{(i)}$,
\begin{subequations} \label{eq:peeling_fit}
\begin{align}
  \{\Xi_j\} &\simeq \xi^{(0)} + \xi^{(1)} R^{-1} + \cdots + \xi^{(p)} R^{-p}, \\
  \{\Upsilon_j\} &\simeq \zeta^{(0)} + \zeta^{(1)} R^{-1} + \cdots + \zeta^{(p)} R^{-p},
\end{align}
\end{subequations}
truncated at some finite order $p$. Comparing the right-hand sides
of Eqs.~\eqref{eq:peeling_fit} and Eq.~\eqref{eq:peeling}, we can
see that if the unwanted effects in the data are all captured by
the subleading terms of the polynomial, then the leading-order terms $\xi^{(0)}$
and $\zeta^{(0)}$ are the asymptotic data. Thus we have,
\begin{equation}
  M^{n-3} \, \Psi_n^{0\, (\ell,m)}(u_i) = \xi^{(0)} + i\zeta^{(0)}.
\end{equation}
For the sake of reducing cumbersome notation, we will refer to the dimensionless
asymptotic Weyl scalars by
\begin{equation}
  \psi_n^{(\ell,m)} \equiv M^{n-3} \, \Psi_n^{0\, (\ell,m)}.
\end{equation}

This fitting procedure is repeated for each $u_i$ to get the full asymptotic
waveform of the $(\ell,m)$ mode, $\psi_n^{(\ell,m)}(u)$.
The extrapolation should be repeatedly performed with increasing extrapolation
order $p$ until $\psi_n^{(\ell,m)}(u)$ converges to a desired tolerance.

In summary, this extrapolation procedure accomplishes
\begin{equation} \label{eq:extrap_summary}
  \lim_{R\rightarrow\infty} R^{5-n} M^{n-3} \, \Psi_n^{(\ell,m)}(u, R) = \psi_n^{(\ell,m)}(u),
\end{equation}
and is readily available in the open-source python module \texttt{scri}~\cite{MobleScri, Boyle2013, Boyle2016, Boyle2014}.
This extrapolation procedure is also used in the same way to find the asymptotic
gravitational wave strain $h^0$ from the finite-radius extracted strain $h$,
\begin{equation} \label{eq:extrap_summary_h}
  \lim_{R\rightarrow\infty} R M^{-1} \, h^{(\ell,m)}(u, R) = h^{0\,(\ell,m)}(u).
\end{equation}

\subsection{Junk Radiation}
\label{sec:junk_radiation}

Binary black hole simulations suffer from a spurious but strong burst of gravitational
radiation that is emitted at the start of the evolution~\cite{HigginBotham2019,Lovelace2009,Varma2018}. This ``junk'' radiation
propagates outward through the domain and should pass through the outer boundary without
affecting the rest of the simulation.

We found that the extracted waveforms for $\Psi_2$, $\Psi_1$, and $\Psi_0$ all
showed effects of the
junk radiation reflecting off the outer boundary and propagating back into
the domain. In addition, there was also significant evidence
that the junk radiation was self-interacting after being initially
emitted and scattering off the nonflat geometry back into the domain before
even reaching the outer
boundary. The radial falloff of the junk radiation is
subleading for $\Psi_{n\ge 3}$ but not for $\Psi_{n \le 2}$. If the junk
radiation is not subleading, then the extrapolation procedure will amplify
its effects rather than removing them.

Three schemes were implemented to reduce the magnitude of the junk radiation, mitigate
the effect of backscattered junk in the waveform data, and prevent the reflection
of junk off the outer boundary. All of these procedures had a negligible
effect on the run time of the simulation.

In order to reduce the junk overall, a different choice of initial
data gauge was made, which was found to lower the magnitude of the
junk by roughly 80\%. The default gauge choice for initial data in SpEC
is the superposed Kerr-Schild (SKS) gauge, which allows one to create initial
data with near-extremal parameters~\cite{Lovelace2008}. Instead, the superposed harmonic Kerr (SHK)
gauge was used~\cite{Varma2018}. This gauge has the benefit of lower junk
radiation at the cost of being unable to create initial data for BBH runs with
high spin. A maximum effective dimensionless spin of around 0.7 can be reached, but
anything higher would require the SKS gauge.

In order to reduce the backscattered junk, we first notice that the
junk radiation is not subleading in radial falloff for $\Psi_{n \le 2}$.
Thus to limit the contribution of junk to the data, we must place the innermost extraction radius
closer to the coordinate
center. We set up 24 extraction
radii, evenly spaced in inverse radius, from $2\lambdabar_0$ to about
$21 \lambdabar_0$, where $\lambdabar_0 = 1 / \omega_{0}$ is the initial
reduced gravitational wavelength as determined by the orbital frequency of the
binary from the initial data.
Since $\Psi_1$ and $\Psi_0$ have such sharp falloff
with radius, the amplitudes of waveforms quickly
fall below the noise floor $\varepsilon$
determined by the simulation resolution. If the extraction radii with
insignificant waveform data are included in the set of data to be extrapolated,
Eq.~\eqref{eq:data_to_extrapolate}, then the extrapolation will not converge.
To improve the extrapolation then, we exclude these insignificant radii
from the extrapolation. For $\Psi_1$ and $\Psi_0$, we
determine the cut-off radius
$R_c$ at each value of retarded time for which we will exclude data from an
extraction radius if it is larger than $R_c$. The value of $R_c$ is defined to be the
radius at which the dominant mode of $\Psi_1$ (or $\Psi_0$) is equal to $\varepsilon$.
For the numerical results in Sec.~\ref{sec:extrap_convergence} we used $\varepsilon=10^{-9}$.

Preventing the junk from reflecting off the outer domain boundary would properly
require improving the
boundary conditions, which is a nontrivial task. Rather than take
this approach, we decided to prevent reflection by effectively ``deleting'' the
outgoing burst when it reached the outer boundary. More specifically, the outer
part of the domain where extraction takes place is constructed from concentric
spherical shells. We
extend the domain with an additional spherical
shell that has no extraction radii and within which the entire burst of junk
radiation will be contained when it reaches the outer boundary. Once the junk is
inside this extra shell we can stop the simulation, delete the extra shell, and
continue the simulation with the now smaller domain. As a rough heuristic, the
burst of junk radiation is typically $\lesssim 450\,M$ wide, so we extend the outer
boundary of the domain by adding an extra $250\,M$-wide spherical shell.
When the peak of the junk radiation reaches the outer boundary, the first half of
  the junk pulse will have already been reflected so that the entire burst of junk
  radiation can be contained within the extra shell. We ensure
that the coordinates inside the domain do not shift when the extra shell is deleted
so that this procedure has no adverse effect on the waveforms being extracted.

\section{Numerical Results}

All numerical work, apart from extrapolation in postprocessing, was done using SpEC. The extrapolation was done with
\texttt{scri}~\cite{MobleScri}.

\subsection{Shifted Kerr}
\label{sec:tetrad_errors}

We begin by testing this extraction-extrapolation procedure with an analytic case in
order to verify convergence of the extrapolation procedure
to the correct result.

For a Kerr spacetime in Kerr-Schild coordinates, the tetrad in Eqs.~\eqref{eq:tetrad}
based on spheres of constant coordinate radius will be neither orthonormal nor
aligned with the principal null directions. The outgoing null tetrad vector points
radially outward from the coordinate center and does not take into account any shift
due to the angular momentum of the spacetime. As expected, these effects are
most pronounced at small radii $r \lesssim 100\,M$.
Furthermore, if the center of mass of the black hole is offset by a distance $\delta z$ along the $z$-axis from the coordinate center, then this further misaligns the outgoing tetrad null vector.

For a Kerr metric in Kerr-Schild coordinates with a center of mass shifted by $\delta z$,
the only nonzero asymptotic Weyl scalar modes are
\begin{subequations}\label{eq:shifted_kerr_weyls}
\begin{align}
  \psi_0^{(2,0)} &\equiv \hspace{4.5mm} \lim_{R\rightarrow\infty} R^5 \Psi_0^{(2,0)} = \sqrt{\frac{24\pi}{5}} \left ( a^2 - 2i a \delta z - \delta z ^2 \right ), \\
  \psi_1^{(1,0)} &\equiv \hspace{4.5mm} \lim_{R\rightarrow\infty} R^4 \Psi_1^{(1,0)} = \sqrt{6\pi} \left ( a i - \delta z \right ), \\
  \psi_2^{(0,0)} &\equiv \hspace{4.5mm} \lim_{R\rightarrow\infty} R^3 \Psi_2^{(0,0)} = - \sqrt{4\pi}, \\
  \psi_3^{(1,0)} &\equiv \hspace{4.5mm} \lim_{R\rightarrow\infty} R^4 \Psi_3^{(1,0)} = \psi_1^{(1,0)}, \\
  \psi_4^{(2,0)} &\equiv \hspace{4.5mm} \lim_{R\rightarrow\infty} R^5 \Psi_4^{(2,0)} = \psi_0^{(2,0)},
\end{align}
\end{subequations}
where $a$ is the Kerr spin parameter, and $M=1$. Usually, the Weyl scalars of a Kerr
spacetime are considered with respect to a Kinnersley tetrad,
in which the only nonzero Weyl scalar is $\Psi_2$; for our tetrad, there is a
nonzero mode for each of the Weyl scalars even when $\delta z = 0$. Since
Kerr is a nonradiating spacetime, notice that
the leading orders for $\Psi_4$ and $\Psi_3$ are $R^{-5}$ and
$R^{-4}$. This demands that the power of $R$ in Eq.~\eqref{eq:extrap_summary}
be adjusted accordingly for extrapolating $\Psi_4$ and $\Psi_3$ in this case.

\begin{figure}[t]
  \centering
  \includegraphics[width=\columnwidth]{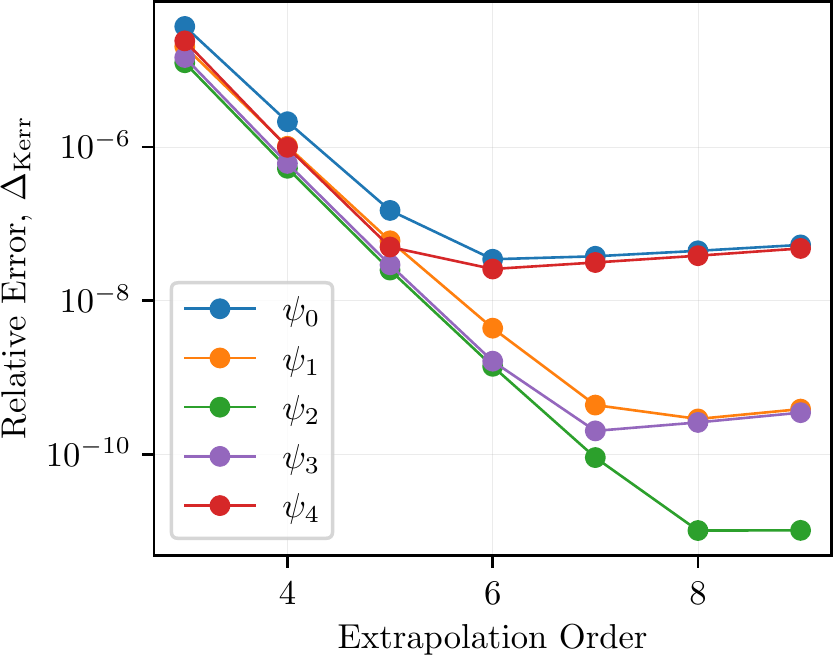}
  \caption{%
    The relative error in the asymptotic Weyl scalars computed from a Kerr spacetime
    in Kerr-Schild coordinates with the center of mass shifted by $\delta z = 1\,M$.}
  \label{fig:kerr_extrap_order}
\end{figure}

Using SpEC, we computed a Kerr spacetime with the
center of mass shifted by $\delta z = 1\,M$. The Weyl scalar mode weights up to
$\ell_\text{max}=8$ were determined at 10 extraction radii equally spaced in
inverse radius from $R_\text{min} = 10\,M$ to $R_\text{max} = 500\,M$. Since the
spacetime is time-independent, we need not worry about the added complication of
choosing a parametrization of null rays $u(t,r)$ for this analysis.

For a range of extrapolation orders, we computed a measure of the relative error
in each computed asymptotic Weyl scalar,
\begin{equation}
  \Delta_\text{\tiny Kerr} = \left | \psi_n^{(\ell_0,m_0)} \right |^{-1} \sqrt{\sum_{\ell,m} \left |\psi_n^{(\ell,m)} - \widehat{\psi}_n^{(\ell,m)} \right |^2},
\end{equation}
where $\widehat{\psi}_n$ denotes the computed asymptotic Weyl scalar, $\psi_n$
denotes the analytic asymptotic Weyl scalar, and $(\ell_0, m_0)$ is the only
nonzero analytic mode. The results for $3 \le p \le 9$ are plotted in
Fig.~\ref{fig:kerr_extrap_order}.

As the extrapolation order increases, the errors decrease exponentially until
they converge. Since we are using 10 extraction radii, we can have a fitting
polynomial of $p<10$. However, using a value of $p \sim p_\text{max}$ will result
in overfitting. This is especially the case for complicated dynamic spacetimes,
as will be discussed in Sec.~\ref{sec:extrap_convergence}. Even with a simple
spacetime like Kerr, the error begins to slowly increase because of overfitting for
$p>6$ with $\psi_4$ and $\psi_0$.

The extrapolation convergence with numerical resolution in the simulation grid
was also investigated. Since SpEC employs a pseudospectral method, we define
the average radial grid density as the number of radial spectral collocation
points divided by the coordinate distance between the outer and inner domain
boundaries. The region inside
the apparent horizon is excised so the excision
surface is the inner boundary of the domain.

\begin{figure}[t]
  \centering
  \includegraphics[width=\columnwidth]{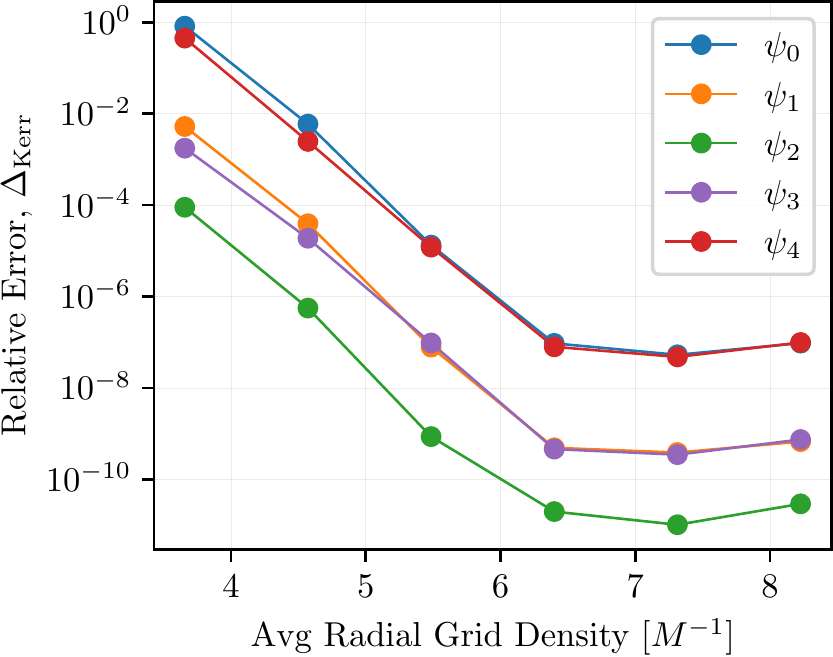}
  \caption{%
    The relative error in the asymptotic Weyl scalar mode weights for a
    Kerr spacetime with the center of mass shifted by $\delta z = 1\,M$. The
    average radial grid density is given by the number of radial spectral
    collocation points divided by the distance between the outer domain boundary
    and the excision boundary.}
  \label{fig:kerr_grid_density}
\end{figure}

Figure~\ref{fig:kerr_grid_density} shows the relative error in the asymptotic
Weyl scalars for $p=9$ as a function of average radial grid density. We see that
the error decreases exponentially as the resolution is increased until the errors
converge.

\subsection{Binary Black Hole Coalescence}

For a complicated dynamical spacetime, like that of a binary black hole coalescence,
we do not have the luxury of comparing the computed asymptotic Weyl scalars to
known analytic values. Instead, we analyze the convergence behavior of the extrapolation
procedure in general. We can also analyze the amount by which the computed
asymptotic Weyl scalars violate the Bianchi identities, which gives us a
self-consistency test against exact general relativity.

\subsubsection{Extrapolation Convergence}
\label{sec:extrap_convergence}

A 20-orbit equal-mass precessing binary black hole inspiral, coalescence, and
ringdown were simulated with dimensionless spins,
\begin{align*}
  \chi_{A} &= \begin{pmatrix} \hphantom{-}0.4684, & 0.1803, & -0.3287\end{pmatrix},\\
  \chi_{B} &= \begin{pmatrix} -0.1924, & 0.0285, & -0.2284\end{pmatrix},
\end{align*}
and 24 extraction radii equally spaced in inverse
radius between $R_\text{min} = 73\,M$ and $R_\text{max} = 770\,M$.

To provide a measure of the convergence of the extrapolation procedure, we compute
the time-averaged relative difference between a waveform $f_p$ found with extrapolation
order $p$ and a waveform $f_{p-1}$ found with extrapolation order $p-1$,
\begin{equation} \label{eq:rel_diff_measure}
  \Delta_{p,p-1} = \frac{1}{u_H - u_0} \int_{u_0}^{u_H} \frac{| f_p(u) - f_{p-1}(u) |}{| f_p(u) |}\,du,
\end{equation}
where $u_0$ is the time of the simulation after the junk radiation has passed and $u_H$ is the
time at which the common horizon forms.

We expect $\Delta_{p,p-1}$ to decrease as $p$ increases as in the case for  the
Kerr spacetime, cf.~Fig~\ref{fig:kerr_extrap_order}. However, with a dynamic
spacetime we have the added complication of choosing an appropriate value for the
retarded time $u$ that accurately parametrizes outgoing null rays. Any
choice of $u$ that poorly parametrizes the null rays will result in errors in the
extrapolation procedure. For the most part, our ansatz for $u(t,r)$, Eqs.~\eqref{eq:corrected_u}, shows a
significant improvement over simply using $u=t-r$. However, there is still room
for future work in
improving the choice of $u$. The net effect is that $\Delta_{p,p-1}$ will decrease
until the extrapolating polynomial begins to fit to artifacts from the
choice of $u$ and numerical noise. Higher extrapolation orders will have a
build up of error and so it will be important to decide on an optimal value for $p$.

\begin{figure}[t]
  \centering
  \includegraphics[width=\columnwidth]{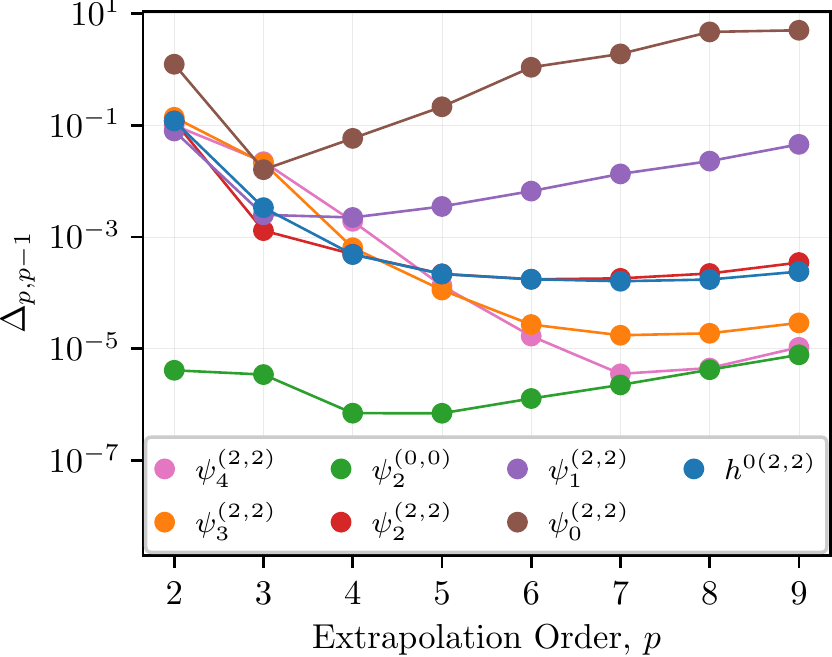}
  \caption{%
    Extrapolation convergence of the extracted Weyl scalars for a precessing
    binary black hole inspiral and merger. The relative difference measure
    $\Delta_{p,p-1}$, Eq.~\eqref{eq:rel_diff_measure}, is plotted for the
    dominant mode of each Weyl scalar. Note that all $p=1$ waveforms are taken to
    be the finite-radius waveforms from the outermost extraction radius. This is
    done to provide a comparison with unextrapolated data.
  }
  \label{fig:bbh_convergence}
\end{figure}

Figure~\ref{fig:bbh_convergence} shows the relative difference of successive
extrapolation orders for each Weyl scalar. The quantities $h$, $\Psi_4$, and $\Psi_3$ show
convergence in the extrapolation of the dominant mode up to $p=7$, after which overfitting errors start
to build up. It appears that the (0,0) mode of $\Psi_2$ does not benefit much from the
extrapolation procedure and is relatively constant with $p$. This permits an extrapolation
order to be chosen that improves the subleading (2,2) mode.

As expected, $\Psi_1$ and $\Psi_0$ are not able to converge to the
same tolerance as the other Weyl scalars with
slower radial falloff. Pleasantly enough, $\Psi_1$ shows some improvement
with extrapolation and converges to about $\mathcal{O}(10^{-3})$.
Before the implementation of the techniques mentioned in Sec.~\ref{sec:junk_radiation},
extrapolation of $\Psi_1$ and $\Psi_0$ was severely unstable even at $p=2$.

As mentioned in the introduction, the most immediate future work resulting from
acquiring asymptotic waveforms is to develop a procedure for completely fixing
the BMS gauge freedom of numerical waveforms. For this purpose, it is specifically
the Weyl scalars $(\Psi_4, \Psi_3, \Psi_2)$ that are of primary importance~\cite{Moreschi1988,Lehner2007,Boyle2016}. Here
we see that for some extrapolation order, we are able to get all three waveforms
respecting the leading falloff to a relative error of $\mathcal{O}(10^{-5})$.

\begin{figure}[t]
  \centering
  \includegraphics[width=\columnwidth]{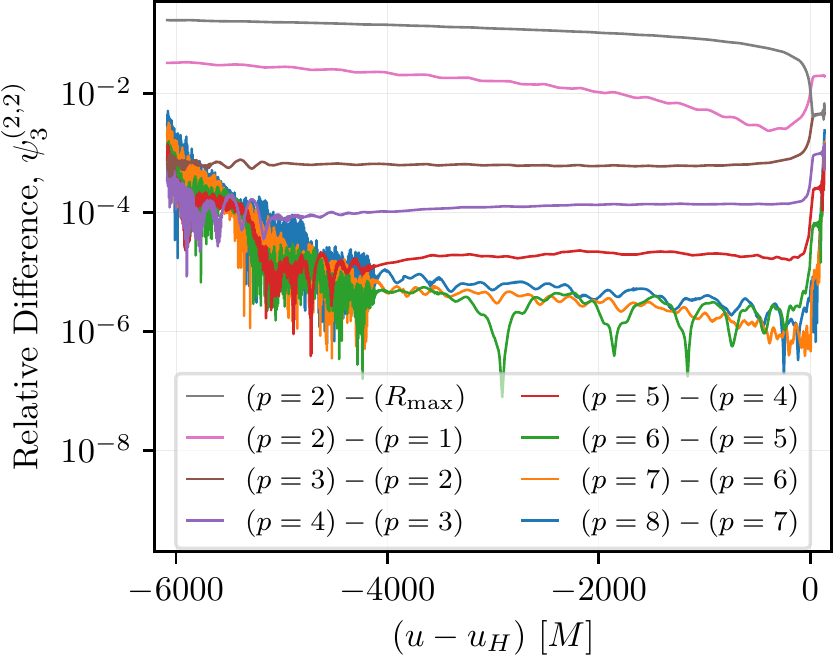}
  \caption{%
    Relative difference of $\psi_3^{(2,2)}$ computed from successive extrapolation
    orders for a precessing binary black hole inspiral, merger, and ringdown.
    The extrapolation order $p=2$ waveform is compared with the unextrapolated
    waveform of the outermost extraction radius. The formation of the common
    horizon occurs at time $u_H$. The time-averaged value of each curve is the
    value of a point for the $\psi_3^{(2,2)}$ curve in Fig.~\ref{fig:bbh_convergence}.
  }
  \label{fig:psi3_convergence}
\end{figure}

Instead of time-averaging the relative difference in a waveform from
two successive extrapolation orders, we can plot the relative difference
as a function of $u$ to see where in the waveform convergence is improving
or diverging. In Fig.~\ref{fig:psi3_convergence}, we have chosen to study the
convergence behavior of the $\psi_3^{(2,2)}$ waveform since it shows both good
convergence behavior for $p\le7$ and a buildup of overfitting errors for $p>7$.

By plotting the full waveform we can see that there is a difference in convergence
behavior for the early inspiral and late inspiral.
The late inspiral converges to a tolerance that is almost two orders of magnitude
lower than the earliest part of the inspiral. This effect is seen with all of the
Weyl scalars. Thus for late inspiral alone, we can expect even better convergence
behavior than shown in Fig.~\ref{fig:bbh_convergence}.
This is to be expected. Near-field effects fall off as $\lambdabar/r$ decreases.
Since $\lambdabar$ decreases
when the binary is closer to merger, so also do the near-field effects even
at a fixed radius. Therefore, the waveform at times closer to merger will be less contaminated
by near-field effects, so it is easier for the extrapolation procedure
to separate the asymptotic waveform from these near-field effects.
Further discussions about the extrapolation procedure can be found in~\cite{Boyle2019,Boyle2009}.

\subsubsection{Bondi Gauge Analysis}
\label{sec:bondi_gauge_analysis}

The Bianchi identities provide a convenient tool to provide a self-consistency
test on asymptotic NR waveforms. In an asymptotic spacetime, Bondi gauge is any
choice of coordinates in which the metric and its derivatives approach Minkowski
spacetime asymptotically. Our extrapolation
procedure assumes an asymptotically flat spacetime,
which should result in Bondi-gauge waveforms on $\mathscr{I}^{+}$.
By taking the Bianchi
identities written in the Newman-Penrose formalism and applying
the assumptions for Bondi gauge, we are
left with a set of constraint equations that must be satisfied for any consistent
set of Bondi gauge waveforms,
\begin{subequations}\label{eq:bondi_constraints}
\begin{align}
  \Psi_4^0 &= - \ddot{h}^0, \label{eq:BI_psi4}\\
  \Psi_3^0 &= \frac{1}{\sqrt{2}} \eth \dot{h}^0, \label{eq:BI_psi3}\\
  \dot{\Psi}_3^0 &= -\frac{1}{\sqrt{2}} \eth \Psi_4^0, \label{eq:BI_dt_psi3}\\
  \dot{\Psi}_2^0 &= -\frac{1}{\sqrt{2}} \eth \Psi_3^0 + \frac{1}{4}\bar{h}^0 \Psi_4^0, \label{eq:BI_dt_psi2}\\
  \dot{\Psi}_1^0 &= -\frac{1}{\sqrt{2}} \eth \Psi_2^0 + \frac{1}{2}\bar{h}^0 \Psi_3^0, \label{eq:BI_dt_psi1}\\
  \dot{\Psi}_0^0 &= -\frac{1}{\sqrt{2}} \eth \Psi_1^0 + \frac{3}{4}\bar{h}^0 \Psi_2^0. \label{eq:BI_dt_psi0}
\end{align}
\end{subequations}
where an overdot signifies a derivative with respect to $u$. We are using
the $\eth$ operator as defined for a spin-weighted function $f$ of spin weight $s$,
\begin{equation}
    \eth f = - \frac{\left ( \sin\theta \right )^{s}}{\sqrt{2}} \left ( \frac{\partial}{\partial\theta} + \frac{i}{\sin\theta} \frac{\partial}{\partial\phi}  \right ) \left [ \left ( \sin\theta \right )^{-s} f \right ],
\label{eq:eth_definition}
\end{equation}
which acts on the spin-weighted spherical harmonics\footnote{The ${}_{-2}Y_{2m}$
  SWSHs as SpEC defines them are given in Eqs.~(C.25--C.27) in~\cite{Boyle2019}.
  } as the ladder-operator,
\begin{subequations}\label{eq:eth_def}
\begin{align}
  \eth\, {}_{s} Y_{\ell m} &= \sqrt{\frac{1}{2} (\ell - s)(\ell + s + 1) }\;{}_{s+1} Y_{\ell m}, \\
  \bar{\eth}\, {}_{s} Y_{\ell m} &= -\sqrt{\frac{1}{2} (\ell + s)(\ell - s + 1) }\;{}_{s-1} Y_{\ell m}.
\end{align}
\end{subequations}
The factors that appear in Eqs.~\eqref{eq:bondi_constraints} may seem to disagree with the
existing literature. See Appendix \ref{app:tetrad_conventions} for a discussion on these differences.
The derivation of Eqs.~(\ref{eq:BI_psi3}--\ref{eq:BI_dt_psi0}) assumes that tetrads have been chosen
such that $h^0=\bar{\sigma}^0$; the Bianchi identities themselves do not depend directly on $h$.
Following the considerations in Sec.~\ref{sec:tetrad_transformations_limit} and
Appendix~\ref{sec:abandon_all_hope_ye_who_enter_here}, the use of $h$ in these equations leaves a possible
subleading term of $\sigma$ unaccounted for. This would mean that these constraint
equations with $h$ are only approximate constraints. Nonetheless, comparisons with CCE waveforms suggest that
this subleading term in $\sigma$ is negligible for our current precision, and even these approximate
constraints are satisfied to a tolerance that allows for practical application.

\begin{figure*}[t]
  \centering
  \includegraphics[width=\textwidth]{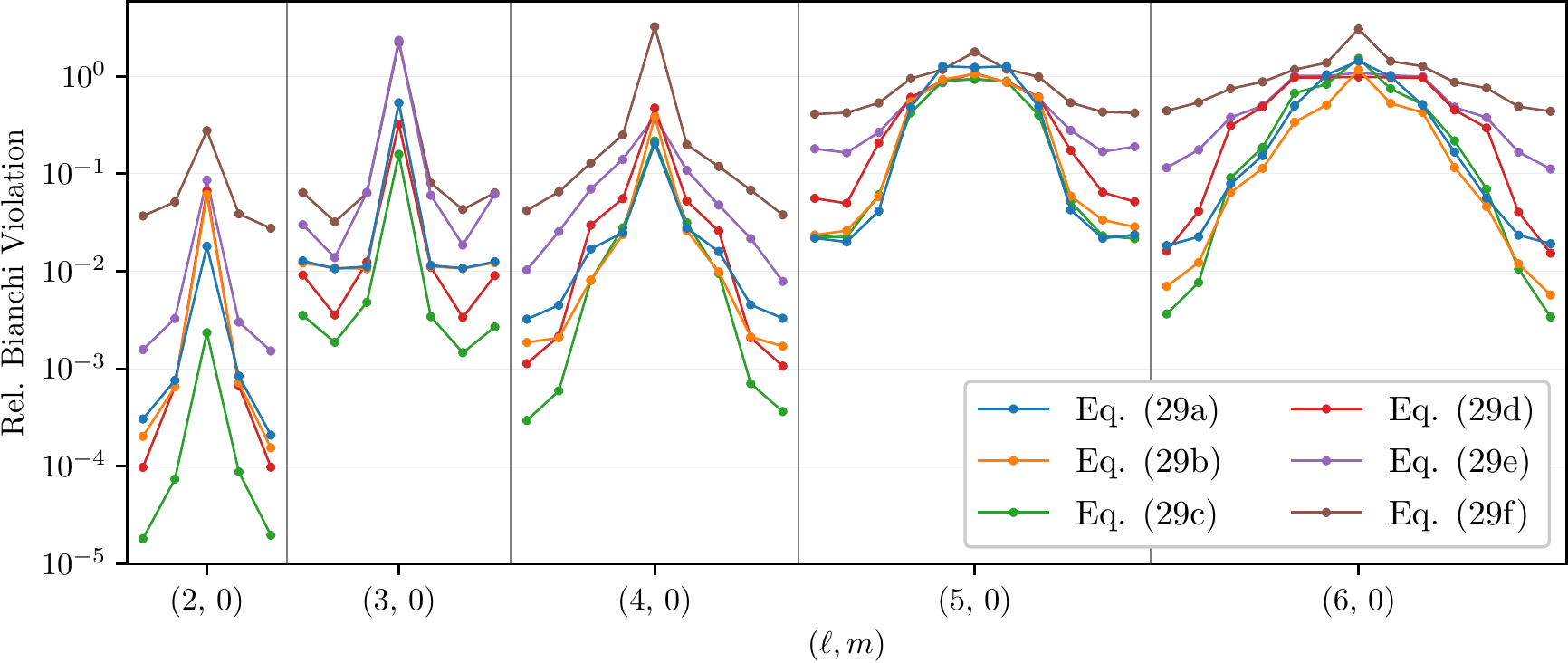}
  \caption{%
    The relative magnitude of the violation of the Bianchi identity constraints,
    Eqs.~\eqref{eq:bondi_constraints}, by numerical Bondi-gauge asymptotic waveforms of
    a binary black hole coalescence. For each value of $\ell$, the modes are plotted
    from $(\ell,-\ell)$ to $(\ell,\ell)$ in order of increasing $m$.
    Each letter in the legend refers to the equation in Eqs.~\eqref{eq:bondi_constraints}
    that is being plotted. Specifically, the values plotted here are the left-hand sides
    of the equations minus the right-hand sides, all scaled by magnitude of the left-hand sides.
    A full discussion of this data is found in Sec.~\ref{sec:bondi_gauge_analysis} after
    Eqs.~\eqref{eq:eth_def}. The modes for each value of $\ell$ have been connected for ease of
    visualization.
  }
  \label{fig:bianchi_violations}
\end{figure*}

Using the information from Fig.~\ref{fig:bbh_convergence} and plots like Fig.~\ref{fig:psi3_convergence}
for each asymptotic quantity, we chose the following values of $p$ for each waveform to test the Bondi gauge
constraints:
\begin{itemize}
  \item $p=7$ for $\psi_4$ and $\psi_3$
  \item $p=5$ for $\psi_2$ and $h^0$
  \item $p=3$ for $\psi_1$
  \item $p=2$ for $\psi_0$
\end{itemize}
Using these waveforms, we can find the relative magnitude of the violations
of Eqs.~\eqref{eq:bondi_constraints}. The deviation from equality is scaled with
respect to the magnitude of the left-hand side of the equation. For each mode,
we take the time average of the violation, setting the initial time to when the initial junk
radiation has passed and setting the final time to $u_H + 80\,M$. The results are
plotted in Fig.~\ref{fig:bianchi_violations}. The time derivatives were performed by fitting a cubic
spline to the waveform and then evaluating the derivative of the spline. Since the
sampling of the data is not uniform in time---with a higher density of points near
merger---we performed a minimization of the violations while varying the density
of the time sampling used in each time derivative.

For the modes that predominantly contribute to the waveform---the $(\ell, \pm\ell)$
and $\left (\ell, \pm (\ell - 1) \right )$ modes---we see violations from Bondi
gauge between $\mathcal{O}(10^{-5})$ and
$\mathcal{O}(10^{-2})$.
The $h^0$ and $\psi_4$ waveforms are of the greatest interest for gravitational
wave astronomy. Although we cannot make any direct statements on how well any individual
waveform satisfies the Bondi constraints, we can parse out some more information
by considering Eqs.~(\ref{eq:BI_psi4}--\ref{eq:BI_dt_psi3}). All three equations
only involve $(h^0, \psi_4, \psi_3)$, and Eq.~\eqref{eq:BI_dt_psi3} is the only
constraint equation that does not include $h^0$. Although Eq.~\eqref{eq:BI_psi3}
and Eq.~\eqref{eq:BI_dt_psi3} are effectively the same relation, just differing by
an overall time derivative, the latter demonstrates smaller violations by roughly
half an order of magnitude. This may imply that a large part of the violation
is due to $h^0$, which would not be unreasonable given that an entirely different
extraction procedure is used for the strain. It has also been observed in
several SXS waveforms that the $h^0$ waveform seems to contain more noise than
the $\psi_4$ waveform. A further analysis of the RWZ extraction procedure for
the strain may shed more light on this.

\section{Conclusion}

All gravitational waveforms have an inherent infinite-dimensional set of gauge
freedoms. When working with asymptotic waveforms at $\mathscr{I}^{+}$, we can
understand transformations between waveforms in different asymptotic coordinates
via the BMS group. Before attempting to build any phenomenological or surrogate
models from NR waveforms, we must both ensure that the waveforms are free from
all near-field effects and also be able to systematically fix the BMS gauge freedom.
This is crucial if we want to separate artifacts
of gauge from the actual physical information in the waveform.

A method for fixing the BMS gauge freedom has been proposed by Moreschi~\cite{Moreschi1987,Moreschi1988}, which
requires reliable extraction of the asymptotic quantities of $h$, $\Psi_4$,
$\Psi_3$, and $\Psi_2$. The extraction procedure implemented in this paper, using
the real characteristic fields of the Weyl tensor evolution equations, is
efficient and readily implementable given the standard 3+1 variables from any
NR code. We have demonstrated a successful implementation in the Spectral
Einstein Code.

The extraction procedure achieves its efficiency at the cost of using a tetrad
choice that is not guaranteed to be orthonormal nor aligned with the principal
null directions of the spacetime. However, we have demonstrated that nonorthonormal
and misaligned tetrads can still be used in getting the asymptotic Weyl
scalars as long as the spurious effects of the tetrad choice fall off with radius at
orders subleading to those specified by the peeling theorem. This paper has explored an
extrapolation procedure by which we can determine the asymptotic waveform data
from the finite-radius extracted data. Using a coordinate-shifted Kerr metric,
we have shown that the extraction and extrapolation procedure is able
to recover the correct asymptotic values. For a precessing, unequal mass ratio,
binary black hole coalescence we have shown that we can find convergence in
the extrapolation procedure for $h$, $\Psi_4$, $\Psi_3$, and $\Psi_2$, while extrapolation still leads to improvement
for $\Psi_1$ and $\Psi_0$. We discussed several methods to reduce the effect of junk
radiation in waveforms resulting from binary black hole initial data.

There are several limiting factors to the extrapolation procedure. As ansatzes,
we have taken the choice of conformal scaling function
Eq.~\eqref{eq:conformal_scaling_function}, the expansion of the Weyl scalars as
a polynomial Eq.~\eqref{eq:peeling},
and the approximate parametrization of null rays Eq.~\eqref{eq:corrected_u}.
An improvement in any one of these may improve the extrapolation convergence.
Despite these limitations, we are able to obtain numerical waveforms for the full
set of Weyl scalars that agree with those of an asymptotic Bondi-gauge spacetime
up to a relative error of $\mathcal{O}(10^{-2})$ for the first few dominant modes.
For the waveforms specifically required for the BMS gauge-fixing procedure, we
are able to obtain waveforms that agree with asymptotic Bondi-gauge waveforms
up to a relative error of $\mathcal{O}(10^{-3})$. Further analysis can be performed
once other extraction procedures, such as CCE, produce asymptotic waveforms to
compare against.

By expanding upon the robust and well established wave extraction method of
SpEC, we have presented the first production-level waveforms for the entire
set of Weyl scalars that are immediately ready for use as tools for gravitational
wave astronomy. Having the full set of Weyl scalars allows us to use the Bianchi
identities to test our extracted waveforms against exact general relativity and
provide hard upper bounds on their accuracy. This analysis is straightforward
to perform and can test each waveform mode individually, as we have
demonstrated in Fig.~\ref{fig:bianchi_violations}. A small public catalog of
simulations with the full set of Weyl scalar waveforms will soon be made available.
The Weyl characteristic field extraction-extrapolation procedure that we have
presented has now set the stage for a reliable method that
will finally provide the gravitational wave astronomy community with completely
gauge-fixed waveforms.

\section{Acknowledgments}

We are grateful to Will Throwe, Eamonn O'Shea, Crist\'obal Armaza, and Gabriel
Bonilla for insightful discussions. Computations were performed with the High
Performance Computing Center and the Wheeler cluster at Caltech. This work was
supported in part by the Sherman Fairchild Foundation and by NSF Grants
No.~PHY-2011961, No.~PHY-2011968, and No.~OAC-1931266 at Caltech and NSF Grants No.~PHY-1912081
and No.~OAC-1931280 at Cornell.

\appendix

\section{Complex Weyl Characteristic Fields}
\label{app:complex_char_fields}

Requiring the Faraday tensor to be divergenceless and
satisfy the Bianchi identity results in two constraint equations and two evolution
equations for the Maxwell electric and magnetic fields. In a similar way,
requiring the Weyl tensor to be divergenceless%
\footnote{
  The divergence of the Weyl tensor is properly sourced by the stress-energy tensor,
  $$\nabla^a C_{abcd} = \nabla_{[d} \left ( -T_{c]b} + \frac{1}{3} T g_{c]b} \right ),$$
  where here $T_{ab}$ is the stress-energy tensor, and thus only vanishes in vacuum,
  cf.~Eq.~\eqref{eq:Weyl_divergence}. This is analogous to the divergence of the
  Faraday tensor vanishing in the absence of sources.
}
and satisfy the Bianchi identities,
\begin{subequations}
\begin{align}
  \nabla_{[a}C_{bc]de} &= 0, \\
  \nabla^a C_{abcd} &= 0, \label{eq:Weyl_divergence}
\end{align}
\end{subequations}
results in two constraint equations and two evolution equations for $E_{ij}$ and
$B_{ij}$. Since we are interested in the propagation of radiation, we will focus
on the evolution equations. Just as in the Maxwell case, we have two coupled
evolution equations for $E_{ij}$ and $B_{ij}$, which we can combine into a single
equation,
\begin{equation}
  \partial_t Q_{ij} - N^k \partial_k Q_{ij} - i N^k \partial_k Q_{l(i} \epsilon_{j)}{}^{kl} = S_Q,
\end{equation}
where $Q_{ij} = E_{ij} + i B_{ij}$, and $S_Q$
is all of the source terms. These source terms are purely algebraic in $E_{ij}$ and
$B_{ij}$. We can further decompose
the quantity $Q_{ij}$ with respect to the geometry of the simulation domain. In the
region we would be extracting $Q_{ij}$, the domain is constructed of concentric
spherical shells, that is, a radial foliation of the spatial hypersurfaces.
If we have a 2-sphere metric $q_{ij}$ and an outgoing spatial radial vector $r^a$,
then the spatial and symmetric tensor $Q_{ij}$ can be decomposed irreducibly into a
scalar function $\mathcal{C}$, a vector $\mathcal{C}_{i}$, and a
transverse-traceless tensor $\mathcal{C}_{ij}$,
\begin{subequations}
\begin{align}
  Q_{ij} &= \mathcal{C} \left ( r_{i} r_{j} - \frac{1}{2} q_{ij} \right ) + 2 r_{(i} \mathcal{C}_{j)} + \mathcal{C}_{ij}, \\
  \mathcal{C} &= \left ( E_{ij} + i B_{ij} \right ) r^i r^j, \\
  \mathcal{C}_i &= \left ( E_{jk} + i B_{jk} \right ) r^j q^{k}{}_{i}, \\
  \mathcal{C}_{ij} &= \left ( E_{kl} + i B_{kl} \right ) \left ( q^{k}{}_{i} q^{l}{}_{j} + \frac{1}{2} q^{kl} q_{ij} \right ).
\end{align}
\end{subequations}
Thus we can express the Weyl scalars in terms of the complex characteristic fields,
\begin{subequations}\label{eq:weyl_from_complex_char_fields}
\begin{align}
  \Psi_4 &= \mathcal{C}_{ij} \bar{m}^i \bar{m}^j, \\
  \Psi_3 &= \frac{1}{\sqrt{2}}\mathcal{C}_{i} \bar{m}^i, \\
  \Psi_2 &= \frac{1}{2}\mathcal{C}, \\
  \Psi_1 &= -\frac{1}{\sqrt{2}}\mathcal{C}_{i} m^i, \\
  \Psi_0 &= \mathcal{C}_{ij} m^i m^j.
\end{align}
\end{subequations}
It is simpler numerically to store and work with real numbers.
Using the following identities~\cite{alcubierre2008},
\begin{subequations}
\begin{align}
 i m_i &= -r^j m^k \epsilon_{ijk},\\
 i \bar{m}_i &= r^j \bar{m}^k \epsilon_{ijk},
\end{align}
\end{subequations}
we can rewrite the three complex fields as the six real
characteristic fields in Eqs.~\ref{eq:characteristic_fields}.

\section{Tetrad Conventions}
\label{app:tetrad_conventions}

The goal of this section is to express the relations between asymptotic quantities
in Bondi gauge in a way that is completely agnostic of sign convention and scale
factors. As such, all the assumptions and results in this section are only valid
with a Minkowski metric. We start by defining a sign variable $s_0$ to account
for different choices of the metric signature,
\begin{equation}
  s_0 = \begin{cases} 1 & \text{for metric signature }(-,+,+,+) \\ -1 & \text{for metric signature }(+,-,-,-) \end{cases}
\end{equation}
For the sake of simplicity, all variables introduced in this section that are
named $s_\text{n}$ will be used to generalize a sign convention and can only take
the value $\pm 1$. In this section, $g_{ab}$ and $\eta_{ab}$ are the $(-,+,+,+)$
signature metrics and explicit factors of $s_0$ will be used to account for metric
signature.

In the literature, it is common to define a complex null tetrad by first constructing $l_a$ to
be a null vector tangent to outgoing null hypersurfaces parametrized by constant retarded time
$u$,
\begin{equation}
  l_a \propto (du)_a.
\end{equation}
The ingoing null tetrad vector $n^a$ is then defined by enforcing the normalization
$l_a n^a = -s_0$. There remains the freedom to introduce a scaling by $\lambda$
that still satisfies the normalization, $(\lambda l_a)(\lambda^{-1} n^a)= -s_0$.
We can absorb this freedom, which includes a sign ambiguity, into the definition
of $l_a$ by defining it as
\begin{equation}
  l_a = -\frac{\lambda}{\sqrt{2}} (dt - dr)_a.
\end{equation}
While $\lambda$ parametrizes the boost freedom of the tetrad, there is still a
spin freedom on the choice of $m^a$, for which we can see that
$m^a \mapsto e^{i\Theta} m^a$ does not affect the normalization $m^a \bar{m}_a = s_0$.
Therefore, we absorb this freedom, parametrized by $0\le\Theta<2\pi$, into the
definition of $m^a$,
\begin{equation}
  m_a = \frac{e^{i\Theta}}{\sqrt{2}} \left ( d\theta + i d\phi \right )_a,
\end{equation}
This orientation is chosen so that for $\Theta=0$, on the $z$ axis we would find
that $(d\theta)_a$ points along the positive $x$ axis. Throughout this section we
are defining $\eth$ as appropriate to each author's definition of $m^a$.

The Christoffel symbols of the second kind contain no factors of $s_0$ so they
are agnostic to metric signature,
\begin{equation}
  \Gamma^c_{ab} = \frac{1}{2} g^{cd} \left ( \partial_b g_{da} + \partial_a g_{db} - \partial_d g_{ab} \right ),
\end{equation}
Using the above definition for the Christoffel symbol, there is a choice of sign
convention on the definition of Riemann tensor, which we parametrize by $s_3$,
\begin{equation}
  s_3 R^a{}_{bcd} = \partial_c \Gamma^a_{db} - \partial_d \Gamma^a_{cb} + \Gamma^a_{ce} \Gamma^e_{db} - \Gamma^a_{de}\Gamma^e_{cb},
\end{equation}
Note that a factor of $s_0$ appears for the lowered-index Riemann tensor,
\begin{equation}
  R_{abcd} = s_3 s_0 \left( \partial_c \Gamma_{adb} - \partial_d \Gamma_{acb} + \Gamma_{ace} \Gamma^e_{db} - \Gamma_{ade}\Gamma^e_{cb} \right ).
\end{equation}
We then need to define the sign variables $s_1$ and $s_2$ to take in account the
choice of sign in the definitions of the Weyl scalars and the Newman-Penrose shear
$\sigma$,
\begin{subequations}
\label{eq:s1_s2_def}
\begin{align}
  \Psi_4 &= s_1 C_{abcd} n^a \bar{m}^b n^c \bar{m}^d, \\
  \sigma &= s_2 m^a m^b \nabla_a l_b,
\end{align}
\end{subequations}
where here the terms on the right-hand sides are in each author's \textit{own}
convention. The Bondi gauge Bianchi identities can now be written as
\begin{subequations}
\begin{align}
  \dot{\Psi}_3^0 &= - \frac{\lambda e^{i\Theta}}{\sqrt{2}} \eth \Psi_4^0, \\
  \dot{\Psi}_2^0 &= - \frac{\lambda e^{i\Theta}}{\sqrt{2}} \eth \Psi_3^0 + \frac{1}{\sqrt{2}} s_0 s_2 \lambda \sigma^0 \Psi_4^0, \\
  \dot{\Psi}_1^0 &= - \frac{\lambda e^{i\Theta}}{\sqrt{2}} \eth \Psi_2^0 + \frac{2}{\sqrt{2}} s_0 s_2 \lambda \sigma^0 \Psi_3^0, \\
  \dot{\Psi}_0^0 &= - \frac{\lambda e^{i\Theta}}{\sqrt{2}} \eth \Psi_1^0 + \frac{3}{\sqrt{2}} s_0 s_2 \lambda \sigma^0 \Psi_2^0.
\end{align}
\end{subequations}
A list of the conventions for various papers is given in Table~\ref{tab:conventions}.

We can also define a parameter $\zeta$ to account for different
scaling factors of the gravitational-wave strain,
\begin{equation}
  h = \zeta^{-1} \left [ \frac{1}{2} \left ( h_{\theta\theta} - h_{\phi\phi} \right ) - i h_{\theta\phi} \right ],
\end{equation}
where $h_{ab} = s_0 (g_{ab} - \eta_{ab})$. From this we can write
the relation between $\Psi_4^0$ and $h^0$ as
\begin{align}
  \Psi_4^0 &= - \left ( s_1 s_3 \zeta \lambda^{-2} e^{-2i\Theta} \right ) \ddot{h}^0,
\end{align}

In order to convert a quantity in the SpEC convention to a different convention,
the appropriate factors can be determined by
\begin{subequations}
\begin{align}
  \Psi^{0\,[\text{X}]}_n &= s_0 s_1 s_3 \left ( \lambda e^{i\Theta} \right )^{2-n} \Psi^{0\,[\text{SpEC}]}_n, \\
  h^{0\,[\text{X}]} &= s_0 \zeta^{-1} e^{-2i\Theta} h^{0\,[\text{SpEC}]},
\end{align}
\end{subequations}
where all of the parameters are from the column of convention [X] in the table.
Although we can easily relate $h^0$ and $\Psi_4^0$ between different conventions,
the situation is far more complicated for $\sigma^0$. These complexities will
be discussed in Appendix~\ref{sec:abandon_all_hope_ye_who_enter_here}.

\begin{table}[t]
  \centering
  \renewcommand{\arraystretch}{1.2}
  \begin{tabular}{@{}c@{\hspace*{5mm}}c@{\hspace*{5mm}}r@{\hspace*{5mm}}r@{\hspace*{5mm}}r@{\hspace*{5mm}}r@{\hspace*{5mm}}r@{}}
    \Xhline{3\arrayrulewidth}
     & SpEC & MB & NP & ADLK & BR & C \\
    \hline
    $s_0$     & $1$ & $-1$        & $-1$        & $1$         & $1$  & $-1$        \\
    $s_1$     & $1$ & $1$         & $-1$        & $1$         & $-1$ & $-1$        \\
    $s_2$     & $1$ & $1$         & $1$         & $-1$        & N/A  & $1$         \\
    $s_3$     & $1$ & $1$         & $1$         & $1$         & $1$  & $1$         \\
    $\lambda$ & $1$ & $-\sqrt{2}$ & $-\sqrt{2}$ & $-\sqrt{2}$ & $1$  & $-\sqrt{2}$ \\
    $\Theta$  & $0$ & $0$         & $\pi$       & $0$         & $0$  & $0$         \\
    $\zeta$   & $1$ & $2$         & N/A         & $1$         & $1$  & N/A         \\
    reference  & \cite{Boyle2019}      & \cite{Boyle2016,Lehner2007} &  \cite{Newman1968} & \cite{Ashtekar2019} & \cite{Bishop2016} & \cite{chandrasekhar1992} \\
    \Xhline{3\arrayrulewidth}
  \end{tabular}
  \caption{
    Sign conventions and scaling factors for various papers. For convenience, a shorthand name for each
    convention is given in the first row. N/A signifies that the particular convention is not specified in that paper.
  }
  \label{tab:conventions}
\end{table}

\section{Subleading tetrad hazards}
\label{sec:abandon_all_hope_ye_who_enter_here}

In Sec.~\ref{sec:tetrad_transformations_limit}, we discussed that in the asymptotic
  limit the Weyl scalars and the strain $h$ are invariant under tetrad transformations that leave the leading
  order tetrad behavior unchanged. However, this does not hold for all the Newman-Penrose
  scalars. Most importantly it does not hold for the shear $\sigma$. Although we are not extracting
  $\sigma$ from simulations, the analysis of numerical waveforms using the BMS group still requires understanding
  how $\sigma$ relates to the Weyl scalars and $h$. Furthermore, a formidable difficulty arises when
  attempting to establish a connection with the literature. Waveform quantities cannot be generally converted
  between the different formalisms because the subleading tetrad behavior is often not specified sufficiently.
  This appendix will explore the effects of subleading tetrad behavior on the asymptotic
quantities that are of primary interest to the study of gravitational radiation.

The waveform quantities that fall off as $1/r^2$ or faster are vulnerable to
dependence on the definitions of the tetrads \emph{off} $\mathscr{I}^+$.
The reason for these corrections is simple, but calculating them is laborious and establishing complete
  agreement of the competing conventions is deeply vexing, especially because
  the subleading behavior in $1/r$ of tetrads is not nearly so universally
  prescribed as the leading behavior described in Appendix \ref{app:tetrad_conventions}.

 The discussion of this appendix is closely related to the concept of tetrad
 rotations, which is covered at length in previous publications~\cite{Newman:1962cia, Campanelli:1998jv}.
 The challenge that we face here, however, distinguishes itself because the
  distinct calculations, especially those performed during Cauchy evolution,
  often do not guarantee that the tetrad basis is orthonormal or null in the
  bulk of the spacetime, only at its boundary.
 Therefore, the alteration between conventions is somewhat more free even than
  generic $\mathcal{O}(1/r)$ tetrad rotations.

To motivate this discussion, first consider the havoc generated by the
  simple alteration of the angular tetrads at subleading order in $1/r$ (holding
  for this illustrative sketch $l^\prime = l$ and $n^\prime = n$),
  \begin{equation}
    m^{\prime \alpha} = m^\alpha + \frac{1}{r}\left( A m^\alpha + B \bar m^\alpha\right).
  \end{equation}
 In particular, the Newman-Penrose spin coefficient
 $\sigma = m^{\alpha} m^{\beta} \nabla_{\alpha} l_{\beta} \sim \mathcal{O}(r^{-2})$ is altered as
  \begin{equation}
    \sigma^\prime = \sigma + \frac{1}{r}\left(2 B \rho + 2 A \sigma\right) + \frac{1}{r^2}\left(A^2 \sigma + 2AB \rho + B^2 \bar\sigma\right),
  \end{equation}
where $\rho = \bar{m}^{\alpha} m^{\beta} \nabla_{\alpha} l_{\beta}$, and
  $\rho = \bar{\rho}$ by assumption.
  However, because $\rho\sim \mathcal{O}(1/r)$, we find that the definition
  of $\sigma$ that we had hoped to standardize now depends on the
  \emph{subleading} values of the tetrad $m^\alpha$. Fortunately, in this
  restricted case, none of the leading contributions to the Weyl scalars are
  altered, but it is not hard to construct alterations to $l$ and $n$ at
  subleading order in $1/r$ that would cause disruption all down the chain
  of Weyl scalars according to the peeling theorem.

In Appendix Sec.~\ref{sec:subleading_tetrad_cce},
  we derive the alteration between the SpEC
  tetrad and the tetrad used in SXS CCE \cite{Moxon:2020gha}. In Appendix Sec.
\ref{sec:tet_rot_cce} we expand the correction between the SXS CCE tetrad and a generic
  asymptotic null tetrad, which is an easier comparison to perform because we
  can take advantage of the properties of null tetrad rotations.
 In each case, we propagate the tetrad alteration to determine the final
  modifications to the asymptotic values of the waveform quantities
  $h, \sigma, \Psi_4, \Psi_3, \Psi_2, \Psi_1$, and $\Psi_0$.

In each of these sections, we denote the tetrads of the various conventions with
text subscripts or superscripts (e.g. $l_{\text{SpEC}}$ for the
 SpEC tetrad convention). To determine the subleading dependence of the tetrads
in the different conventions, it is often necessary to expand the tetrads in
powers of inverse $r$, which we denote with the order of inverse $r$ in
parentheses (e.g. $l_{\text{CCE}}^{(0)}$). Implicitly, this is written
  as the $r$ coordinate in the SpEC convention, but because we only work in a
  limited expansion in powers of inverse $r$, all of the statements would be
  unchanged if working in the Bondi-Sachs $\mathring r$. To avoid confusion,
we do not use the superscript $0$ as in the body of the text to denote the
leading contribution to a waveform quantity asymptotically. Instead, we use the
  explicit power of $r$ explicitly, writing for instance $\Psi_4^0$ as
  $\Psi_4^{(1)}$.

\subsection{Subleading tetrads in CCE}
\label{sec:subleading_tetrad_cce}

In this section, we use the tetrads for a CCE formalism described in
  \cite{Moxon:2020gha},
  {\allowdisplaybreaks
  \begin{subequations} \label{eq:CCE_tetrad}
  \begin{align}
    m^\mu_{\text{CCE}}
    &= -\frac{1}{\sqrt{2} r} \left(\sqrt{\frac{K + 1}{2}}q^\mu
      - \sqrt{\frac{1}{2(1 + K)} J} \bar q^\mu\right),\\
    n^\mu_{\text{CCE}}
    &= \sqrt{2} e^{-2\beta} \bigg(\delta^\mu{}_u
      - \frac{1}{2}\left(1 + r W\right)\delta^\mu{}_r \notag\\&\hspace{1.9cm}+ \frac{1}{2} \bar U q^\mu
      + \frac{1}{2} U \bar q^\mu\bigg),\\
    l^\mu_{\text{CCE}} &= \frac{1}{\sqrt{2}} \delta^\mu{}_r,
  \end{align}
\end{subequations}}%
where the Bondi-Sachs scalars $J, K, \beta, V, U$ and coordinates are as defined
in \cite{Moxon:2020gha}, which each represent components of the metric in
  Bondi-Sachs coordinates.
  We assume that the tetrads constructed in Eq.~\eqref{eq:tetrad} for the
  SpEC Cauchy simulation are in agreement with the CCE tetrad,
  Eqs.~(\ref{eq:CCE_tetrad}), asymptotically.
We expand the relation in powers of inverse $r$, denoting (dropping the ``CCE''
for brevity, understanding that the order subscripts in this appendix section
will apply exclusively to the tetrad derived in the CCE formalism),
\begin{subequations}
  \begin{align}
    m^\mu_{\text{CCE}} &= m^\mu_{(0)} + \frac{1}{r} m^\mu_{(1)} + \mathcal{O}(r^{-2}),\\
    l^\mu_{\text{CCE}} &= l^\mu_{(0)} + \frac{1}{r} l^\mu_{(1)} + \mathcal{O}(r^{-2}),\\
    n^\mu_{\text{CCE}} &= n^\mu_{(0)} + \frac{1}{r} n^\mu_{(1)} + \mathcal{O}(r^{-2}).
  \end{align}
\end{subequations}

Importantly, when attempting to compare the tetrads in the disparate
  coordinate systems, we need to be aware of the alterations associated with the
  conversion between the Bondi-Sachs coordinate system and the Cauchy
  coordinates.
  For simplicity of the current presentation, we expand the Bondi-Sachs coordinates in
  terms of the Cauchy radial coordinate,
  \begin{subequations}
    \begin{align}
      \mathring{u} &= u + \frac{1}{r} \mathring{u}^{(1)} + \mathcal{O}(r^{-2}), \\
      \mathring{r} &= r + \mathring r^{(0)} + \frac{1}{r} \mathring{r}^{(1)} + \mathcal{O}(r^{-2}), \label{eq:CCE_coord_r}\\
      \mathring{x}^A &= x^A + \frac{1}{r} \mathring{x}^{(1) A} + \mathcal{O}(r^{-2}),
    \end{align}
  \end{subequations}
 Unfortunately, the differences between these coordinate systems depend on the
  myriad choices in constructing a CCE evolution associated with a particular
  Cauchy evolution, including extraction surface and data on the initial
  hypersurface.
 The quantities $\mathring u^{(1)}$, $\mathring r^{(1)}$, and
  $\mathring x^{(1) A}$ can be determined numerically for a particular Cauchy
  and CCE evolution, but practical implementations of that
  calculation is beyond the scope of the current discussion.

  A contribution $r^{(0)}$ to Eq.~\eqref{eq:CCE_coord_r}, should it be
  nonvanishing, has even more dire consequences on attempts to establish
  asymptotic correspondence. This coordinate alteration impacts the leading tetrads,
  \begin{subequations}
  \begin{align}
    m_{(0)}^\mu = m^\mu_{\text{SpEC}} + l^\mu m^\nu_{\text{SpEC}} \partial_\nu \mathring r^{(0)}, \\
    n_{(0)}^\mu = n^\mu_{\text{SpEC}} + l^\mu n^\nu_{\text{SpEC}} \partial_\nu \mathring r^{(0)}.
  \end{align}
\end{subequations}
Fortunately, such impact is easy to notice, as it will manifest as a
nonvanishing inner product
$n^\mu_{\text{SpEC}} g_{\mu \nu} n^\nu_{\text{SpEC}}$ or
$n^\mu_{\text{SpEC}} g_{\mu \nu} m^{\nu}_{\text{SpEC}}$ at $\mathscr{I}^+$.
We will assume from here on in this appendix that any such pathology has been
avoided in the Cauchy code and that we may safely set
$\mathring r^{(0)} = 0$.
Therefore, the leading tetrads at $\mathscr{I}^+$ are assumed to be in agreement
between the Cauchy code and CCE: $l^\mu_{(0)} = l^\mu_{\text{SpEC}}$,
$m^\mu_{(0)} = m^\mu_{\text{SpEC}}$, $n^\mu_{(0)} = n^\mu_{\text{SpEC}}$.

Given the agreement between asymptotic tetrads, we can use the subleading
  metric contracted with the leading tetrads (from either formalism) to infer
  the metric components that act as inputs to the tetrad definitions
  in Eqs.~\eqref{eq:CCE_tetrad}. In particular,
  \begin{subequations}
  \begin{align}
    J^{(1)} &= g_{m m}^{(1)},\\
    U^{(2)} &= g_{n m}^{(1)},\\
    \beta^{(1)} &= \frac{1}{2}  g_{n l}^{(1)},\\
    W^{(2)}  &= -\frac{1}{2} g_{n n}^{(1)}.
  \end{align}
\end{subequations}

When all effects are taken into account, the subleading tetrad expressions
for the CCE formalism are
{\allowdisplaybreaks
\begin{subequations} \label{eq:CCE_tetrad_corrections}
\begin{align}
      m_{(1)}^\mu &=
      - J^{(1)} \bar m^\mu_{(0)},\\
      n_{(1)}^\mu &= \sqrt{2} \partial_u \mathring{u}^{(1)} (n_{(0)}^\mu + l_{(0)}^\mu) + \sqrt{2} \partial_u \mathring r^{(1)} l^\mu_{(0)}\notag\\
                  &\hspace{3mm}+ \frac{1}{\sqrt{2}} \partial_u \mathring x^{(1) A} \left(m^{(0)}_A  \bar m_{(0)}^\mu + \bar m^{(0)}_{A}  m_{(0)}^\mu\right) \notag\\
                  &\hspace{3mm} - 2 \sqrt{2} \beta^{(1)} n^{\mu}_{(0)} - W^{(2)} l^\mu_{(0)} \notag\\
                  &\hspace{3mm}- \bar U^{(2)} m_{(0)}^\mu - U^{(2)}\bar m_{(0)}^\mu,\\
                  &\equiv n^{(1)}_n l^\mu_{(0)} + n^{(1)}_l n^\mu_{(0)} + n^{(1)}_m \bar m^\mu_{(0)} + n^{(1)}_{\bar m} m^\mu_{(0)},\\
      l_{(1)}^\mu &= 0.
\end{align}
\end{subequations}}%
To summarize, the subleading contributions to the $m$ tetrad vector amounts simply to a
 $\mathcal{O}(r^{-1})$ rotation in the $m$, $\bar m$ plane, the $n$ tetrad vector has
contributions along all of the original tetrad directions, and the subleading
$l$ tetrad vector vanishes.
We emphasize that the tetrad corrections between the CCE and SpEC tetrads are
not simply a tetrad rotation, so we must consider carefully the effects on the
waveform quantities.

Given the above CCE tetrad, Eqs.~\eqref{eq:CCE_tetrad}, with explicit
$\mathcal{O}(r^{-1})$ parts, Eqs.~\eqref{eq:CCE_tetrad_corrections}, the
conversion between the waveform quantities
  derived from the SpEC tetrad and CCE tetrad are as follows,
  \begin{subequations}
    \begin{align}
      h^{(1) \text{CCE}} &= h^{(1) \text{SpEC}},\\
      \sigma^{(2)\text{CCE}} &= \sigma^{(2) \text{SpEC}}\notag\\&\hspace{.5cm}
      - J^{(1)}\left(\rho^{(1)\text{SpEC}} + \bar \rho^{(1)\text{SpEC}}\right),\\
      \Psi_4^{(1)\text{CCE}} &= \Psi_4^{(1)\text{SpEC}},\\
      \Psi_3^{(2)\text{CCE}} &= \Psi_3^{(2)\text{SpEC}},\\
      \Psi_2^{(3)\text{CCE}} &= \Psi_2^{(3)\text{SpEC}},\\% - \bar\Psi_3 \bar m^{(1)}_l + \Psi_3 m^{(1)}_l \\
      \Psi_1^{(4)\text{CCE}} &= \Psi_1^{(4)\text{SpEC}},\\% + \bar\Psi_2 m^{(1)}_l\\
      \Psi_0^{(5)\text{CCE}} &= \Psi_0^{(5)\text{SpEC}}.% + 2 \Psi_1 m^{(1)}_l + \bar \Psi_2 (m^{(1)}_l)^2
    \end{align}
  \end{subequations}
  The main take-away from this calculation is that the leading strain and all of the Weyl scalars agree between the two formalisms. The shear, however, is a bit of a sticking point. In particular, if the $\rho$ mimics typical Kerr behavior asymptotically and $\sigma^{\text{SpEC}}$ is asymptotically equal to the leading part of the strain, the $\sigma^{\text{CCE}}$ will differ from $\sigma^{\text{SpEC}}$ by an overall sign change.

\subsection{Tetrad rotations between CCE and other formulations}
\label{sec:tet_rot_cce}

Many of the methods of choosing a Newman-Penrose construction at $\mathscr{I}^+$
do not completely specify the tetrad behavior at subleading order.
In this section, we will make some fairly general assumptions about the
construction of the subleading tetrad contribution, and derive the corrections
between possible choices in those constructions.

First, let us consider the case of unrestricted null tetrad rotations.
The condition that the tetrads remain orthonormal and null constrains the set of
degrees of freedom,
{\allowdisplaybreaks \begin{subequations} \label{eq:generic_rot}
\begin{align}
  \bar m^{\mu}_{\text{G}} &= \bar m^{\mu}_{\text{CCE}} + \frac{1}{r} \bigg(-\bar m^{\text{G} (1)}_{l} n^\mu_{\text{CCE}} - \bar m^{\text (1) G}_{n} l^\mu_{\text{CCE}} \notag\\&\hspace{2.4cm}+ \bar m^{\text (1) G}_m \bar m^\mu_{\text{CCE}}\bigg) + \mathcal{O}(r^{-2}),\\
  m^{\mu}_{\text{G}} &= m^{\mu}_{\text{CCE}} + \frac{1}{r} \bigg(- m^{\text{G} (1)}_{l} n^\mu_{\text{CCE}} - m^{\text (1) G}_{n} l^\mu_{\text{CCE}} \notag\\&\hspace{2.4cm}+ \bar m^{\text (1) G}_m  m^\mu_{\text{CCE}}\bigg)+ \mathcal{O}(r^{-2}), \\
  n^\mu_{\text G} &= n^{\mu}_{\text{CCE}} + \frac{1}{r} \bigg(- n^{\text{G} (1)}_{l} n^\mu_{\text{CCE}} - \bar m^{\text (1) G}_n m^\mu_{\text{CCE}} \notag\\&\hspace{2.4cm}- m^{\text{G} (1) }_n \bar m^\mu_{\text{CCE}}\bigg)+ \mathcal{O}(r^{-2}), \\
    l^\mu_{\text G} &= l^{\mu}_{\text{CCE}} + \frac{1}{r} \bigg( n^{\text{G} (1)}_{l} l^\mu_{\text{CCE}} - \bar m^{\text (1) G}_l m^\mu_{\text{CCE}} \notag\\&\hspace{2.4cm}- m^{\text{G} (1) }_l \bar m^\mu_{\text{CCE}}\bigg)+ \mathcal{O}(r^{-2}).
\end{align}
\end{subequations}}

Generic rotations of the form Eq.~\eqref{eq:generic_rot} give rise to numerous
corrections to the waveform quantities,
{\allowdisplaybreaks \begin{subequations} \label{eq:generic_transform_waveform}
  \begin{align}
    &h^{(1)\text G} = h^{(1)\text{CCE}}, \\
    &\Psi_4^{(1)\text{G}} = \Psi_4^{(1)\text{CCE}}, \\
    &\Psi_3^{(2)\text{G}} = \Psi_3^{(2)\text{CCE}} - \Psi_4^{(1)\text{CCE}} m^{\text (1) G}_l, \\
    &\Psi_2^{(3)\text{G}} =\notag\\&\hspace{.5cm} \Psi_2^{(3)\text{CCE}} - 4 \Psi_3^{(2)\text{CCE}} m^{\text (1) G}_l + 2 \Psi_4^{(1)\text{CCE}} \left(m_l^{\text (1) G}\right)^2, \\
    &\Psi_1^{(4)\text G} = \notag\\&\hspace{.5cm} \Psi_1^{(4)\text{CCE}} - 2 \Psi_2^{(3)\text{CCE}} m^{\text (1) G}_l + \bar \Psi_3^{(2)\text{CCE}} \bar m^{\text (1) G}_l m^{\text (1) G}_l \notag\\
&\hspace{.5cm}- \bar \Psi_4^{(1)\text{CCE}} \left(\bar m^{\text (1) G}_l\right)^2 m^{\text (1) G}_l + 3 \Psi_3^{(2)\text{CCE}} \left(m^{\text (1) G}_l\right)^2\notag\\
&\hspace{.5cm}-\Psi_4^{(1)\text{CCE}} \left(m^{\text (1) G}_l\right)^3, \\
&    \Psi_0^{(5)\text G} =\notag\\&\hspace{.5cm} \Psi_0^{(5)\text{CCE}} - 4 \Psi_1^{(4)\text{CCE}} m^{\text (1) G}_l + 5 \Psi_2^{(3)} \left(m^{\text (1) G}_l\right)^2 \notag\\
                &\hspace{.5cm} + \bar \Psi_4^{(1)\text{CCE}} \left(\bar m_l^{\text (1) G}\right)^2 \left(m_l^{\text (1) G}\right)^2 \notag\\
                &\hspace{.5cm}- 4 \Psi_3^{(2)\text {CCE}} \left(m_l^{\text (1) G}\right)^2  + \Psi_4^{(1)} \left(m_l^{\text (1) G}\right)^4, \\
&    \sigma^{(2)\text G} =\notag\\&\hspace{.5cm} \sigma^{(2)\text{CCE}} - 2 \beta^{(1)} m^{\text (1) G}_l - \kappa^{(1)} m^{\text (1) G}_n - \tau^{(1)} m^{\text (1) G}_l \notag\\&\hspace{.5cm}+ m_{\text G}^{(0)\mu} \nabla_\mu m^{\text (1) G}_l - m^{\text (1) G}_l n_{\text G}^{(0)\mu} \nabla_\mu m^{\text (1) G}_l.
  \end{align}
\end{subequations}}%
In Eqs.~\eqref{eq:generic_transform_waveform}, we use the
  standard Newman-Penrose spin coefficient notation
  $\beta=1/2 (n^\alpha m^\beta \nabla_\beta l_\alpha -
  \bar m^\alpha m^\beta \nabla_\beta m_\alpha)$,
  $\kappa=-m^\alpha l^\beta \nabla_\beta l_\alpha$,
  and $\tau=-m^\alpha n^\beta \nabla_\beta l_\alpha$.
This causes significant difficulty in comparing the results from different
formalisms.
However, there is a clear pattern associated with which parts
of the tetrad rotation are important for the waveform comparisons.

In particular, if we merely impose that the null tetrad of the formulation we are
comparing with the CCE results shares an $l$ tetrad vector, we find that the comparison
expressions, Eqs.~\eqref{eq:generic_transform_waveform}, simplifies greatly (denoting as
LPG an ``$l$-preserving generic'' formalism that preserves the CCE $l$ vector -- i.e. $l^\mu_{\text{CCE}} = l^\mu_{\text{LPG}}$),
{\allowdisplaybreaks \begin{subequations}
  \begin{align}
    h^{(1) \text{LPG}} &= h^{(1) \text{CCE}}, \\
    \sigma^{(2) \text{LPG}} &= \sigma^{(2) \text{CCE}}- \kappa^{(1)} m^{\text{LPG} (1)}_n, \\
    \Psi_4^{(1) \text{LPG}} &= \Psi_4^{(1) \text{CCE}}, \\
    \Psi_3^{(2) \text{LPG}} &= \Psi_3^{(2) \text{CCE}}, \\
    \Psi_2^{(3) \text{LPG}} &= \Psi_2^{(3) \text{CCE}},\\
    \Psi_1^{(4) \text{LPG}} &= \Psi_1^{(4) \text{CCE}},\\
    \Psi_0^{(5) \text{LPG}} &= \Psi_0^{(5) \text{CCE}}.
  \end{align}
\end{subequations}}%
We emphasize that in the case where the $l$ tetrad vector is preserved between
formalisms, \emph{all} of the waveform quantities can be directly compared except
for the shear $\sigma$.
In particular, the relationship between the shear and the strain can differ
between formalisms with different subleading tetrads, even if the tetrads
evaluated at $\mathscr{I}^+$ are identical.

% \section{Convention Check (Not for Paper)}
%
% Moreschi-Boyle:
% \begin{align*}
%   h &= \bar{\sigma} \\
%   \psi_4 &= - \ddot{\bar{\sigma}} \\
%   \psi_3 &= - \eth \dot{\bar{\sigma}} \\
%   \dot{\psi_2} &= \eth \psi_3 + \sigma \psi_4
% \end{align*}
%
% ADLK:
% \begin{align*}
%   h &= 2\bar{\sigma} \\
%   \psi_4 &= - \ddot{\bar{\sigma}} \\
%   \psi_3 &= - \eth \dot{\bar{\sigma}} \\
%   \dot{\psi_2} &= \eth \psi_3 + \sigma \psi_4
% \end{align*}
%
% NP:
% \begin{align*}
%   \psi_4 &= - \ddot{\bar{\sigma}} \\
%   \psi_3 &= \eth \dot{\bar{\sigma}} \\
%   \dot{\psi_2} &= -\eth \psi_3 + \sigma \psi_4
% \end{align*}
%
% Chandrasekhar:
% \begin{align*}
%   \psi_4 &= \ddot{h} \\
%   \dot{\psi_2} &= \eth \psi_3 + \sigma \psi_4
% \end{align*}

\bibliographystyle{apsrev4-1}
\bibliography{main}

\end{document}